\documentclass[a4paper,fleqn,usenatbib]{mnras}
\usepackage[english]{babel}
\usepackage[utf8x]{inputenc} 
\usepackage{newtxtext,newtxmath}
\usepackage{longtable}
\usepackage{setspace}
\usepackage{booktabs}
\usepackage{wallpaper}
\usepackage{epstopdf}
\usepackage{eso-pic}

\usepackage[singlelinecheck=off]{caption}
\usepackage{longtable}

\usepackage[T1]{fontenc}
\usepackage{ae,aecompl}

\usepackage{graphicx}	
\usepackage{amsmath}	
\usepackage{amssymb}	

\def\HII{H{\sc ii} }
\def\NN{N$_{2}$H$^{+}$}
\def\15N{$^{15}$NNH$^{+}$}
\def\N15{N$^{15}$NH$^{+}$}

\title[N-fractionation: IRAS 05358+3543]{First interferometric study of enhanced N-fractionation in \NN: the high-mass star-forming region IRAS 05358+3543}

\author[L. Colzi et al.]{L. Colzi$^{1,2}$\thanks{E-mail: colzi@arcetri.astro.it},
F. Fontani$^{2}$,
P. Caselli$^{3}$,
S. Leurini$^{4}$,
L. Bizzocchi$^{3}$,
and G. Quaia$^{1}$
\\
$^{1}$Università degli studi di Firenze, Dipartimento di fisica e Astronomia, Via Sansone 1, 50019 Sesto Fiorentino, Italy\\
$^{2}$INAF-Osservatorio Astrofisico di Arcetri, Largo E. Fermi 5, I-50125, Florence, Italy \\
$^{3}$Max-Planck-Instit\"{u}t f\"{u}r extraterrestrische Physik, Giessenbachstrasse 1, D-85748, Garching bei M\"{u}nchen, Germany\\
$^{4}$INAF – Osservatorio Astronomico di Cagliari, Via della Scienza 5,
Selargius CA 09047, Italy\\
}

\date{Accepted 2019 March 14. Received 2019 March 14; in original form 2019 January 25.}

\pubyear{2018}

\begin{document}
\label{firstpage}
\pagerange{\pageref{firstpage}--\pageref{lastpage}}
\maketitle

\begin{abstract}
Nitrogen (N) fractionation is used as a tool to search for a link between the chemical history of the Solar System and star-forming regions. A large variation of $^{14}$N/$^{15}$N is observed towards different astrophysical sources, and current chemical models cannot reproduce it. With the advent of high angular resolution radiotelescopes it is now possible to search for N-fractionation at core scales.
We present IRAM NOEMA observations of the J=1--0 transition of \NN, \15N and \N15 towards the high-mass protocluster IRAS 05358+3543.
We find $^{14}$N/$^{15}$N ratios that span from $\sim$100 up to $\sim$220 and these values are lower or equal than those observed with single-dish observations towards the same source. Since N-fractionation changes across the studied region, this means that it is regulated by local environmental effects. We find also the possibility, for one of the four cores defined in the protocluster, to have a more abundant \15N with respect to \N15. This is another indication that current chemical models may be missing chemical reactions or may not take into account other mechanisms, like photodissociation or grain surface chemistry, that could be important.

\end{abstract}




 \section{Introduction}
 \label{intro}
The origin of the Solar System is still a highly debated topic. Understanding the origin of the molecules in different Solar System bodies is important to have information on the proto-Solar nebula (PSN), from which our Sun was born. In particular, it is not clear if molecules in pristine Solar System materials, like comets or asteroids, were inherited from the cold and dense PSN, or if they are the result of chemical processing within the Solar protoplanetary disc.
 
One of the most used approaches involves the measurement of the $^{14}$N/$^{15}$N ratio from different molecules. The ratio measured for the PSN from the Solar wind is 441$\pm$6 (Marty et al.~\citeyear{marty10}), and this value is about two times larger than that measured in the terrestrial atmosphere (TA), derived from N$_{2}$, $\sim$272 (Marty et al.~\citeyear{marty09}). Colzi et al.~(\citeyear{colzi18b}) derived a new $^{14}$N/$^{15}$N ratio in the local ISM of 375$\pm$60, broadly consistent with the PSN value and consistent with the values predicted by the Galactic chemical evolution model of Romano et al.~(\citeyear{romano17}).
The PSN value is also larger than that measured in some comets ($\sim$140, observed in CN, HCN and NH$_{2}$, average value from Hily-Blant et al.~\citeyear{hily-blant17} and reference therein) and in carbonaceous chondrites ($\sim$50-250, e.g. van Kooten et al.~\citeyear{vankooten17}). van Kooten et al.~(\citeyear{vankooten17}) found also a different $^{15}$N-enrichment between the amine groups and the nitrile functional groups in the lithic clasts of the Isheyevo carbonaceous chondrite of their study.
New measurements also confirm the trend of low $^{14}$N/$^{15}$N values in pristine Solar System materials, in fact Yang et al.~(\citeyear{yang17}) derived a $^{14}$N/$^{15}$N=130$\pm$15 from CN and $^{14}$N/$^{15}$N=140$\pm$28 from NH$_{2}$ in the outbursting comet C/2015 ER61. Moreover, McGeoch et al.~(\citeyear{mcgeoch18}) determined, with a study of polymers of amino acids, in Allende and Acfer 086 meteorites $^{14}$N/$^{15}$N of $\sim$200 and $\sim$140, respectively.

The relation between the $^{15}$N-enrichments in pristine Solar System material and the natal core is still uncertain, and probably these differences are also coming from different nitrogen reservoirs. $^{14}$N/$^{15}$N ratios were measured through observations towards both low- and high-mass star-forming regions from different molecules: HCN and HNC (e.g. Hily-Blant et al.~\citeyear{hily-blant13}, Zeng et al.~\citeyear{zeng17}, Colzi et al.~\citeyear{colzi18a}, \citeyear{colzi18b}, Magalhães et al.~\citeyear{magalhaes18}), \NN (e.g. Daniel et al.~\citeyear{daniel16}) and NH$_{3}$ (e.g. Gerin et al.~\citeyear{gerin09}). None of these values are consistent with the anti-fractionation ($^{14}$N/$^{15}$N>500) observed with \NN\:by Bizzocchi et al.~(\citeyear{bizzocchi13}), Redaelli et al.~(\citeyear{redaelli18}) and Fontani et al.~(\citeyear{fontani15}). Moreover, a difference between nitrile molecules (i.e. CN-bearing, such as HCN/HNC) and hydrogenated N-bearing molecules (e.g. NH$_{3}$ and \NN) has been claimed, e.g. by Hily-Blant et al.~(\citeyear{hily-blant13}) and Wampfler et al.~(\citeyear{wampfler14}). A different behaviour between HCN/HNC and N$_{2}$H$^{+}$ is also apparent in massive star-forming regions. In fact, Fontani et al.~(\citeyear{fontani15}) found a larger dispersion of values from \NN (from 180 up to 1300), with respect to the values found by Colzi et al.~(\citeyear{colzi18a}) from HCN and HNC (from 250 up to 650) in the same sample of sources in which were also performed other astrochemical studies (e.g. Colzi et al.~\citeyear{colzi18a}, Mininni et al.~\citeyear{mininni18} and Fontani et al.~\citeyear{fontani18}).

A summary of the measured $^{14}$N/$^{15}$N ratios from different molecules can be found in Wirström et al.~(\citeyear{wirstrom16}).

Guzmán et al.~(\citeyear{guzman17}) have measured the $^{14}$N/$^{15}$N from HCN towards 6 protoplanetary discs, and found that most of them shows $^{14}$N/$^{15}$N consistent with comet-like values (80-160, with an average value of 111$\pm$9). However, Hily-Blant et al.~(\citeyear{hily-blant17}) measured a CN/C$^{15}$N ratio of 323$\pm$30 in the disk orbiting the nearby young star TW Hya, and they proposed the CN as a possible present-day reservoir of nitrogen in the Solar neighbourhood.

From a theoretical point of view, two possible mechanisms were proposed to explain $^{15}$N-fractionation: isotope-exchange reactions, favoured at low temperatures (T<20K, e.g. Roueff et al.~\citeyear{roueff15}, Wirström \& Charnley \citeyear{wirstrom18}, Loison et al.~\citeyear{loison19}), or selective photo-dissociation of $^{14}$N$^{15}$N over $^{14}$N$_{2}$ (Heays et al.~\citeyear{heays14}), favoured in the external layers of the discs exposed directly to the illumination of the central (proto-)star. Visser et al.~(\citeyear{visser18}) confirmed that isotope-selective photodissociation could be crucial to understand the N-fractionation in protoplanetary discs, especially those exposed to strong irradiation fields. In particular, for \NN, this is true if the dominant fractionation mechanism are reactions with atomic N.

In this work we will focus on \NN. Terzieva \& Herbst (\citeyear{terzieva00}) proposed that the reactions that cause most of $^{15}$N-enrichment in \NN are:
\begin{subequations}
\begin{gather}
\label{NN-reaction1}{\rm N}_{2}{\rm H}^{+} + {\rm^{15}N} \leftrightarrow {\rm N}^{15}{\rm NH}^{+} + {\rm N}, \\
\label{NN-reaction2} {\rm N}_{2}{\rm H}^{+} + {\rm^{15}N} \leftrightarrow {\rm ^{15}N}{\rm NH}^{+} + {\rm N}.
\end{gather}
\end{subequations}
However, the most recent chemical models have challenged this scenario, due to the recent discovery that reactions \eqref{NN-reaction1} and \eqref{NN-reaction2} do not occur in cold environments due to the presence of an entrance barrier (Roueff et al.~\citeyear{roueff15}, Wirström \& Charnley \citeyear{wirstrom18}). These chemical models fail to reproduce both the anti-fractionation (i.e. $^{14}$N/$^{15}$N>500) measured in cold prestellar cores (Redaelli et al.~\citeyear{redaelli18}) and the large spread of values measured from different high-mass star forming regions (Fontani et al.~\citeyear{fontani15}). This also suggests that chemical reactions may still be missing in existing models, or that the $^{15}$N-enrichment (or the contrary, i.e. $^{15}$N-depletion, which causes anti-fractionation) is a local phenomenon occurring in spatial regions smaller than the beam size of single-dish observations described above. In fact, for example with the single dish observations made by Fontani et al.~(\citeyear{fontani15}), the isotopic ratios obtained were average values over angular sizes of $\sim$30'', which hence include both the large-scale envelope and the dense core(s) embedded within it. A more recent chemical model of Furuya \& Aikawa (\citeyear{furuya18}) proposed that anti-fractionation could be explained introducing other mechanisms in chemical models, like isotope selective photodissociation of N$_{2}$ and grain surface chemistry. We will discuss this point in Sect.~\ref{discussion-corescale}.

In this work we report, for the first time, an interferometric analysis of the isotopic ratio $^{14}$N/$^{15}$N towards the high-mass star-forming protocluster IRAS 05358+3543 (hereafter 05358) from \NN. We discuss the possible different N-fractionation between \15N and \N15 and the possible $^{15}$N-enrichment at core scales ($\sim$5''). N-fractionation in \NN towards this source was already studied by Fontani et al.~(\citeyear{fontani15}), who analysed IRAM-30m observations towards a sample of 26 high-mass star-forming regions, including 05358. In particular, for this source they found that the $^{14}$N/$^{15}$N is lower ($\sim$200) than in the other massive star-forming regions.

The source and the observations are described in Sect.~\ref{observations}, the results are presented in Sect.~\ref{results} and a discussion of the results is presented in Sect.~\ref{discussion}.
 
 \section{Source and observations}
 \label{observations}
 
 05358 is part of a sample of 69 high-mass protostellar objects studied by Sridharan et al.~(\citeyear{sridharan02}), Beuther et al.~(\citeyear{beuther02b}), (\citeyear{beuther02c}) and (\citeyear{beuther02d}) in the millimeter wavelengths. The source lies at a distance of 1.8 kpc in the Auriga molecular cloud complex (Heyer et al.~\citeyear{heyer96}), in the Perseus spiral arm of the Milky Way, and has a bolometric luminosity of 6300 L$_{\odot}$ (Snell et al.~\citeyear{snell90}). The source is part of a complex group of \HII regions: SH 235 is the brightest of four optical \HII regions and it is excited by the O9.5 star BD+351201 (Georgelin \citeyear{georgelin75}). The source 05358 is associated with the \HII region SH 233 which is approximately 25$\arcmin$ west and 5$\arcmin$ south of SH 235.
 
As evidence for massive star-formation, the source is associated with maser emissions (e.g. Minier et al.~\citeyear{minier00}, Hu et al.~\citeyear{hu16})  and massive outflows (e.g. Beuther et al.~\citeyear{beuther02a}, Ginsburg et al.~\citeyear{ginsburg09}). 
Beuther et al.~(\citeyear{beuther07}) resolved 05358 in at least three continuum sub-sources:\\
- mm1: it has a typical spectrum of a young massive protostar where the central source has already started hydrogen burning. The source could be a B1 Zero-Age-Main-Sequence star, with a luminosity of $\sim$10$^{3.72}$ L$_{\odot}$ and a stellar mass of $\sim$13 M$_{\odot}$ (Lang \citeyear{lang92}). From the vicinity of the source, a collimated outflow is ejected with a rate of 6$\times$10$^{-4}$ M$_{\odot}$/yr (Beuther et al.~\citeyear{beuther02a}), indicating that the protostar is still accreting gas. Moreover, the source is associated with a 8.3 GHz emission (VLA data) that is likely from a hypercompact \HII region. The main (sub)mm continuum source, which is at the center of two molecular outflows (Beuther et al.~\citeyear{beuther02a}), is resolved in two separate (sub)mm continuum peaks, mm1a and mm1b; \\ 
-mm2: it consists of at least two sub-sources, mm2a and mm2b, and the position of mm2b shifts from 1.2 mm to 875 $\mu$m. There is also the presence of other peaks (mm2c and mm2d). The source mm2a is detected as a compact one and this could indicate that only this source is a star-forming region, while the others could be transient sources caused by the multiple outflow system in the region;\\
-mm3: for this source there are neither cm nor compact line detection (Leurini et al.~\citeyear{leurini07}) and then it could be a very cold massive core in an early evolutionary stage. The grey-body function used to fit the SED is constrained by the high-frequency ($\sim$700 GHz) measurement by Beuther et al.~(\citeyear{beuther07}) to dust temperatures below 20 K.\\
Table \ref{coordinates} contains the source coordinates.\\

\begin{table}
\begin{center}
\caption{Coordinates of the dust continuum sources as observed by Beuther et al.~(\citeyear{beuther07}).}
 \begin{tabular}{lcccc}
  \hline
  Source  & RA[J2000] & DEC[J2000] \\ 
          & (h m s)& ($\degr$ $\arcmin$ $\arcsec$)  \\
  \hline
 mm1a & 05:39:13.08 & 35:45:51.3\\
 mm1b & 05:39:13.13 & 35:45:50.8\\
 mm2 & 05:39:12.76 & 35:45:51.3\\
 mm3 & 05:39:12.50  &35:45:54.9\\
 mm4$^{a}$ & 05:39:12.46 & 35:45:40.9\\
\hline
  \normalsize
  \label{coordinates}
  \end{tabular}
  \end{center}
  \begin{flushleft}
 \footnotesize
 $^{a}$ Source that is discovered in this work and it was not studied by Beuther et al.~(\citeyear{beuther07}).\\
        \end{flushleft}
       \normalsize
 \end{table}
 
 We carried on observations with the IRAM NOEMA Interferometer of the J=1--0 transition of \NN, \15N and \N15 towards 05358. Rest frequencies are 93.1734 GHz (Cazzoli et al.~\citeyear{cazzoli12}), 90.2638 GHz and 91.2057 GHz (Dore et al.~\citeyear{dore09}), for the J=1--0 transition of \NN, \15N and \N15, respectively. In Table \ref{frequencies} the energy of the upper levels, the Einstein coefficients and rest frequencies of each hyperfine transition are given, except for \NN(1--0) for which only the rotational frequency is given, since the hyperfine structure is not resolved in our spectra. To discuss our results, in Sect.~\ref{discussion-map} we will also use the averaged emission map of the J=5--4, K=0,1 transitions of CH$_{3}$CN. Rest frequencies are 91.987 GHz and 91.985 GHz, respectively (Cazzoli \& Puzzarini \citeyear{cazzoli06}). Because these lines were in the same spectral band of the J=1--0 transition of \NN, synthesized beam, spectral resolution and rms are the same.
\begin{table}
\begin{center}
\caption{Molecular transitions of the J=1--0 transition of \NN, \15N and \N15.}
 \begin{tabular}{lcccc}
  \hline
  Molecule  & Frequency & Transition & $E_{\rm U}$ & log$_{10}$({\it A}$_{\rm ij}^{a}$)  \\ 
          & (GHz) & &(K) &\\
  \hline
 \NN$^{b}$ & 93.1734 & J=1--0 & 4.5 & -4.44034\\
 \hline
 \15N & 90.2635 & J=1--0, F=1--1 & 4.3 & -4.48168\\
 & 90.2639 & J=1--0, F=2--1& & \\
 & 90.2645 & J=1--0, F=0--1 & & \\
 \hline
 \N15 & 91.2043& J=1--0, F=1--1 & 4.4 & -4.46810\\
 & 91.2060 & J=1--0, F=2--1& & \\
 & 91.2086 & J=1--0, F=0--1 & & \\
\hline
  \normalsize
  \label{frequencies}
  \end{tabular}
  \end{center}
   \begin{flushleft}
 \footnotesize
 $^{a}$ Einstein coefficient of the transition;\\
 $^{b}$ For this molecule only the rotational transition between level $i$ and $j$ is given, since with our spectral setup we were not able to resolve the hyperfine structure.
        \end{flushleft}
       \normalsize
  \end{table}
  
Observations were carried out in 4 days from Sep. 29 to Nov. 3,
2016, with 8 antennas in C and D configurations, providing baselines
in between 15 and 240 m, corresponding to an angular resolution
of $\sim3.1\times 3\arcsec$.
The amount of precipitable water vapour was generally in between
5 and 10 mm. Visibility amplitudes and phases were calibrated
on 0552+398 and 0548+378. The absolute flux density scale was
calibrated on MWC349 and LKHA101. The bandpass calibration was
performed on 3C84 or 3C454.3.
To incorporate short spacings to the interferometric maps, we
have reduced and analysed IRAM-30m observations of the J=1--0 transition of \NN, \15N and \N15. These observations were carried out on March 5, 6
and 7, 2017. We have obtained large-scale maps on an angular region
of $\sim120\arcsec$, i.e. about twice the NOEMA primary beam at the
frequency of the \NN(1--0) transition.
The data were obtained with the on-the-fly mapping mode. Pointing
was checked every hour on nearby quasars. Focus was checked at
start of the observations, and after sunset.
Creation of the synthetic visibilities and merging of the two
data sets were performed through standard procedures available in
the GILDAS software package MAPPING.
  
Calibration and imaging were performed using CLIC and MAPPING software of the GILDAS package\footnote{The GILDAS
software is available at http://www.iram.fr/IRAMFR/GILDAS}. The synthesized beam, the final velocity resolution and the rms (root mean square) of the datacubes are given in Table \ref{beam}. 
The analysis of the data was done with MADCUBA\footnote{Madrid Data Cube Analysis on ImageJ is a software developed in the
Center of Astrobiology (Madrid, INTA-CSIC) to visualise and analyse
single spectra and datacubes (Martín et al., in prep., Rivilla et al.~\citeyear{rivilla16}, Rivilla et al.~\citeyear{rivilla17}).} software package. For the analysis we have used the
spectroscopic parameters from the CDMS
molecular catalog\footnote{https://www.astro.uni-koeln.de/cdms/} (Müller et al.~\citeyear{muller01}, \citeyear{muller05}; Endres et al.~\citeyear{endres16}).
\begin{table}
\begin{center}
\caption{Observational parameters.}
 \begin{tabular}{lcccc}
 \hline
 \hline
  Spw & Synthesized beam & PA & $\Delta$v & rms  \\ 
        &  ($\arcsec \times \arcsec$)&(°)& (km s$^{-1}$) &(mJy/beam) \\
  \hline
 Continuum & 3.01 $\times$ 2.61 & -110.18 & -- & 0.2\\
 \NN &  3.04 $\times$ 2.64 & -109.5 & 6.45 & 0.5 \\
 \15N &  3.12 $\times$ 3.08 &-41.91&  0.5 & 7 \\
 \N15 &  3.12 $\times$ 3.08 &-41.91&  0.5 & 7 \\
 \hline
 \normalsize
  \label{beam}
  \end{tabular}
  \end{center}
  \end{table}

 \section{Results}
 \label{results}
 \subsection{Continuum map}
\begin{figure}
\centering
\includegraphics[width=20pc]{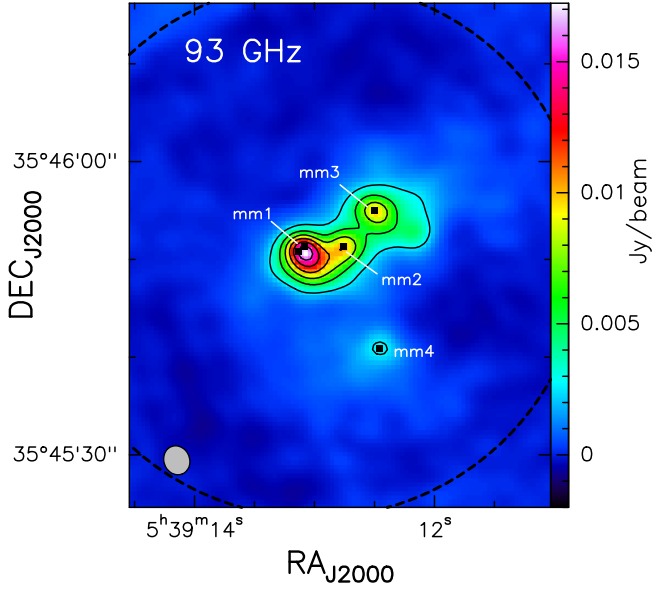}
\caption{93 GHz continuum emission obtained with IRAM NOEMA. The first contour level is the 15\% of the maximum value of the map, that corresponds to $\sim$17 mJy/beam. The other contour levels are the 30\%, 45\%, 60\%, 75\% and 90\% of the maximum value, respectively. The black squares indicate the position of mm1a, mm1b, mm2, mm3 and mm4 as given in the text and in Table \ref{coordinates}. The dashed circle represents the NOEMA field of view and the synthesized beam is the ellipse indicated in the lower left corner.}
\centering
\label{cont-map}
\end{figure}

 In Fig.~\ref{cont-map} the 3 mm continuum map towards the source is shown. With this map we are able to distinguish the already known continuum millimeter sources mm1, mm2 and mm3 (Beuther et al.~\citeyear{beuther07}). However, we are not able to resolve mm1 as a binary system composed by mm1a and mm1b (Table \ref{coordinates}) because of the insufficient spatial resolution (Table \ref{beam}). Moreover, we have observed the continuum emission of a new millimeter source (mm4) at right ascension $\alpha_{\rm 2000}$= 05$^{\rm h}$:39$^{\rm m}$:12.46$^{\rm s}$ and declination $\delta_{\rm 2000}$=35°:45$\arcmin$:40.9$\arcsec$. Since this source is out of the primary beam at 1 mm, it was not analysed by Beuther et al.~(\citeyear{beuther07}) that focused their work in characterising the properties of 05358 at the center of the field of view.
 
 \subsection{Morphology of \NN and $^{15}$N-isotopologues emission}
 \label{molecule-map}
\begin{figure}
\centering
\includegraphics[width=20pc]{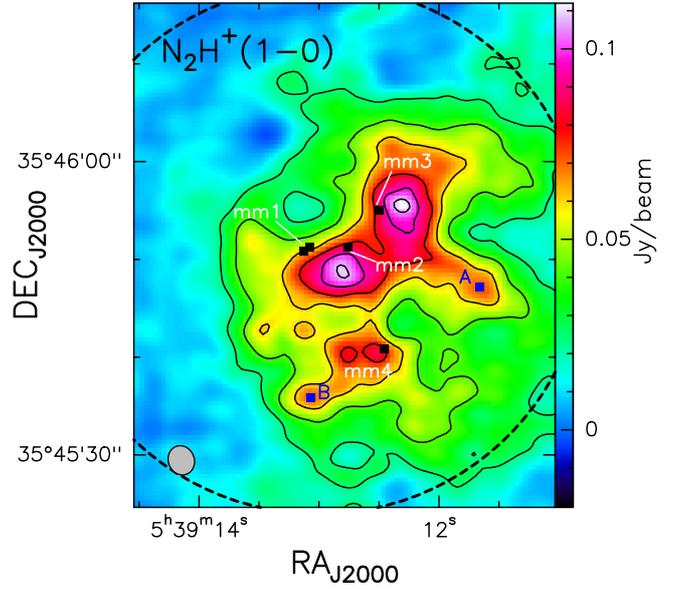}
\caption{Averaged map of \NN(1--0) at 93.1734 GHz (velocity range between -34.6 and -0.1 km s$^{-1}$) obtained from the Widex correlator. The contour levels are 4, 7, 10, 13, 16 and 18 times the 1$\sigma$ rms of the map, equal to $\sim$6 mJy/beam km s$^{-1}$. The black squares indicate the position of mm1a, mm1b, mm2, mm3 and mm4 as given in the text and in Table \ref{coordinates}. The blue squares correspond to the \NN(1--0) emission peaks, A and B, described in the text.
The dashed circle represents the NOEMA field of view and the synthesized beam is the ellipse indicated in the lower left corner.}
\centering
\label{n2hp-map}
\end{figure}

In Fig.~\ref{n2hp-map} we show the averaged emission map of \NN(1--0), which arises mainly from 3 cores: one associated with both mm1 and mm2 not resolved, another associated with mm3, and finally one associated with mm4. We have also highlighted the presence of two other \NN(1--0) emission regions (A and B) and we will discuss them in Sect.~\ref{discussion-map}. It can also be noted that the \NN(1--0) emission peaks are shifted by $\sim$2--3$\arcsec$ with respect to the mm1, mm2 and mm3 continuum sources.  We will discuss this result in Sect.~\ref{discussion-map}.

\begin{figure*}
\centering
\includegraphics[width=43pc]{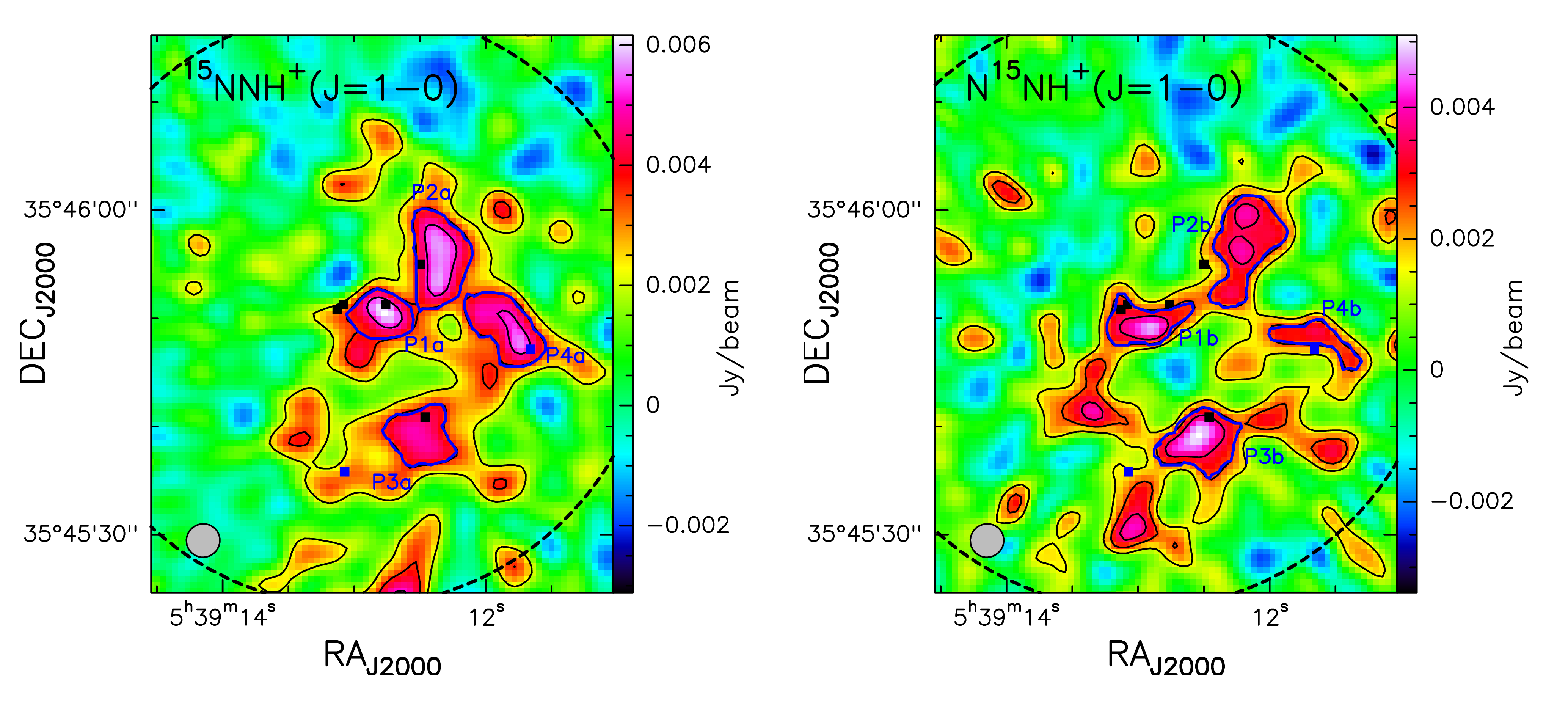}
\caption{\emph{Left panel}: Averaged map of \15N(1--0) at 90.26 GHz (velocity range between -19.3 and -14.2 km s$^{-1}$) obtained from the Narrow correlator. The black contour levels are 3, 5 and 7 times the 1$\sigma$ rms of the map, equal to $\sim$0.5 mJy/beam km s$^{-1}$. \emph{Right panel}: Averaged map of the J=1--0, F=2-1 transition of \N15  (main hyperfine component) at 91.20 GHz (velocity range between  -19.5 and -15 km s$^{-1}$) obtained from the Narrow correlator. The black contour levels are 3, 5 and 7 times the 1$\sigma$ rms of the map, equal to $\sim$0.72 mJy/beam km s$^{-1}$.
In both panels, the blue contours correspond to the 5$\sigma$ level of the averaged maps, from which the spectra have been extracted: P1a, P2a, P3a and P4a from \15N and P1b, P2b, P3b and P4b from \N15. The black and blue squares indicate the positions of the continuum sources and the \NN peak positions (A and B), respectively, as in Fig.~\ref{n2hp-map}. The dashed circle represents the NOEMA field of view and the synthesized beam is the ellipse indicated in the lower left corner.}
\centering
\label{15n-map}
\end{figure*}

In Fig.~\ref{15n-map} the averaged emission maps of the \15N(1--0) and \N15 (1--0) transitions are shown. They are obtained by integrating the lines over the channels with signal, above the 3$\sigma$ level, in the Narrow spectra. In particular, for \N15(1--0) we were able to resolve the hyperfine structure in three distinct components (see Table \ref{frequencies}). However, since the J=1--0, F=1-1 and J=1--0, F=0-1 transitions are near the noise level, we have decided to create the emission map integrating only over the channels that correspond to strongest component, namely the J=1-0, F=2-1 transition. Moreover, in Fig.~\ref{15n-map} the blue contours correspond to the polygons from which we have extracted the spectra, in order to derive the total column densities of the three molecules, and the corresponding $^{14}$N/$^{15}$N ratios, in eight different regions in 05358. The polygons are defined from the 5$\sigma$ of the corresponding averaged map. In particular, P1a, P2a, P3a and P4a correspond to a value of 5$\times$0.72 mJy/beam km s$^{-1}$ (from \15N, left panel of Fig.~\ref{15n-map}), while P1b, P2b, P3b and P4b correspond to a value of 5$\times$0.5 mJy/beam km s$^{-1}$ (from \N15, right panel of Fig.~\ref{15n-map}).

\begin{figure*}
\centering
\includegraphics[width=43pc]{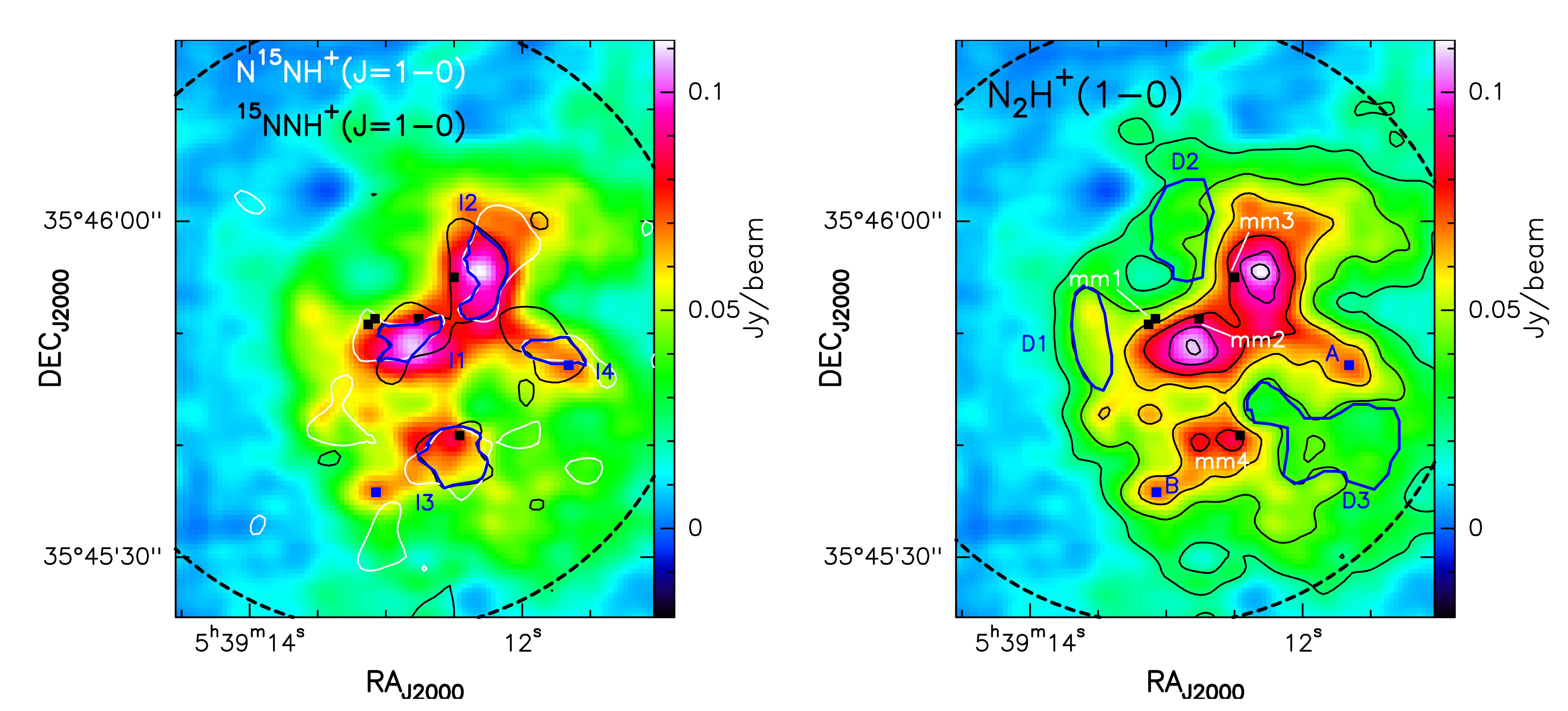}
\caption{\emph{Left panel}: The colours are the averaged emission of \NN(1--0). The black contours are the 5$\sigma$ levels of \15N(1--0) averaged emission and the white contour are the 5$\sigma$ levels of \N15(1--0) averaged emission: in blue the intersection of the black and white contours (I1, I2, I3 and I4).
\emph{Right panel}: The same of Fig.~\ref{n2hp-map}. The blue contours define three different zones D1, D2 and D3, in the diffuse \NN envelope surrounding the dense condensations identified in the left panel and in Fig.\ref{15n-map}.
In both panels, the black and blue squares indicate the positions of the continuum sources and the \NN peak positions (A and B), respectively, as in Fig.~\ref{n2hp-map}. The dashed circle represents the NOEMA field of view.}
\centering
\label{inter-diffuse}
\end{figure*}

We have also extracted spectra from polygons defined as the intersection of the 5$\sigma$ of the two $^{15}$N-isotopologues averaged maps (I1, I2, I3 and I4, left panel of Fig.~\ref{inter-diffuse}). This has been done in order to compare the $^{14}$N/$^{15}$N ratios obtained from \15N and \N15 from the same regions of the source. Finally, we have decided to estimate the $^{14}$N/$^{15}$N ratios in the regions defined in the right panel of Fig.~\ref{inter-diffuse}, D1, D2 and D3, less bright and corresponding to diffuse \NN emission.

\subsection{Fitting procedure and column density calculation}
\label{madcuba}
To fit the lines and to compute the total column densities from the extracted spectra we have used MADCUBA assuming local thermal equilibrium (LTE) conditions. This is a reasonable assumption since the observed cores present volume densities of the order of 10$^{6}$ cm$^{-3}$ (see Beuther et al.~\citeyear{beuther07}), that is higher than the critical densities of the observed lines ($\sim$1.5--2$\times$10$^{5}$ cm$^{-3}$, assuming kinetic temperatures from 20 K up to 50 K).

The MADCUBA-AUTOFIT tool takes into account four parameters to create a synthetic LTE line profile: total column density ({\it N}), excitation temperature ($T_{\rm ex}$), peak velocity ({\it v}), and full width half maximum ($FWHM$). We have fitted all the lines fixing the $T_{\rm ex}$. We have performed the analysis with different $T_{\rm ex}$ values (20, 30, 40 and 50 K) since we had only one transition for each molecules and we were not able to derive it directly from the data. These temperature values are also consistent with the kinetic temperatures derived for 05358 from NH$_{3}$(1,1) by Lu et al.~(\citeyear{lu14}). Leaving free the other parameters ($N$, $v$, and $FWHM$) the AUTOFIT tool compares the LTE synthetic spectra with the observed spectra and provides the best non-linear least squared fit using the Levenberg-Marquardt algorithm. When the algorithm converges, it provides also the associated errors to the parameters. From these parameters, the tool calculates also the line opacity between levels $i$ and $j$, $\tau_{\rm ij}$ (see eq.~(1) and (2) in Rivilla et al.~\citeyear{rivilla18}).

\begin{figure}
\centering
\includegraphics[width=20pc]{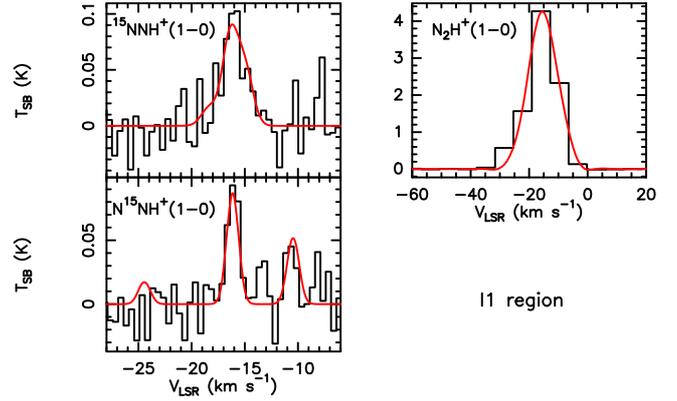}
\caption{Example of spectra extracted from the region I1 (see left panel of Fig.~\ref{inter-diffuse}) in all the detected transitions: \15N(1--0) (top-left), \N15(1--0) (bottom-left) and \NN(1--0) (top-right). For each spectrum the x-axis represents the systemic velocity of the source. The y-axis shows the intensity in synthesised beam temperature units. The red curves are the best fits obtained with MADCUBA (from the method described in Sect.~\ref{madcuba}). Spectra extracted from the other intersecting regions (I2--I4) are shown in Appendix \ref{spectra-inters}.}
\centering
\label{spettriI1}
\end{figure}

The three lines that we have observed, the \NN(1--0), \15N(1--0) and \N15(1--0) transitions, have hyperfine structure. We can resolve it only for \N15(1--0) since the line width found ($\sim$2 km s$^{-1}$) is smaller than the separation in velocity of the hyperfine components (e.g. Fig.~\ref{spettriI1}). However, we have fitted the spectra of both the \15N(1--0) and \N15(1--0) transitions taking the hyperfine structure into account, in order to be consistent with the analysis of the two different $^{15}$N-isotopologues of \NN. Conversely, for \NN(1--0) we have performed the analysis taking into account only the rotational transition J=1--0, as if it was a single line. We did this because the low spectral resolution ($\sim$6.5 km s$^{-1}$) permits the convergence of the fits only fixing a FWHM>6.5 km s$^{-1}$, that is not the real one of the source ($\sim$2 km s$^{-1}$ from the fits of $^{15}$N-isotopologues). The fit to a single transition permits to leave the FWHM as a free parameter and, even if the final FWHM is not reliable, the convergence of the fit is more precise. In Appendix \ref{analysis-nn} a detailed analysis demonstrates that the use of this method gives similar results of the analysis of the same \NN simulated spectrum where the hyperfine structure could be resolved, within the errors.

All the extracted spectra from the P1a--P4a, P1b--P4b, I1, I2, I3 and I4 regions, and the more diffuse \NN emission regions D1, D2 and D3  (see Sect.~\ref{molecule-map}) are shown in Appendix \ref{spectra}. Column densities, FWHM, velocities and opacities derived from the fit of the extracted spectra are given in Appendix \ref{appendix-fit}.

For the main isotopologue, \NN, we have found column densities of the order of 10$^{14}$cm$^{-2}$ in all regions. In particular, column densities towards P1a and P2a are higher than those towards P3a and P4a, of a factor $\sim$1.5. The same trend is followed by P1b and P2b with respect to P3b and P4b, with a decrease of a factor $\sim$3. However, for the $^{15}$N-isotopologues column densities are the same in each region, within the errors, and are of the order of 10$^{11}$cm$^{-2}$--10$^{12}$cm$^{-2}$, depending on the assumed T$_{\rm ex}$. In fact, as a general trend for all the analysed molecules, \NN, \15N and \N15, column densities increases with the T$_{\rm ex}$, but their ratios remain the same, within the uncertainties.

\subsubsection{Diffuse regions}
\label{diffuse}
The \15N(1--0) and \N15(1--0) transitions are not detected in any of the diffuse \NN emission regions D1, D2 and D3, except for \15N(1--0) in D2 for which we have obtained a tentative detection (see Figs.~\ref{spectra-diffuse}). We have considered tentative detections the lines where the peak synthesized beam temperature ($T_{\rm SB}^{\rm peak}$) was between 2.5$\sigma$ and 3$\sigma$ ($2.5\sigma \leq T_{\rm SB}^{\rm peak}< 3\sigma$) , where $\sigma$ is the r.m.s noise of the spectrum. For the non detections we have computed upper limits for the column densities with MADCUBA. They have been derived from the formula 3$\sigma\times\Delta$v/$\sqrt{N_{\rm chan}}$, where $N_{\rm chan}$ is the number of channels covered by the line width $\Delta$v. In particular, we have considered as $\Delta$v an average value of the FWHM derived from the same transitions towards P1a--P4a and P1b--P4b (Table \ref{polygon-fit}): $\Delta$v$_{^{15}{\rm NNH}^{+}}=2.3\pm 0.2$ km s$^{-1}$ and $\Delta$v$_{{\rm N}^{15}{\rm NH}^{+}}=1.7\pm 0.4$ km s$^{-1}$.
The upper limits and tentative detection fit results from diffuse \NN emission regions are reported in Table \ref{diffuse-fit}.

\subsubsection{Extracted spectra from IRAM-30m beam}
\label{extract-30m}
In order to compare our results with those obtained by Fontani et al.~(\citeyear{fontani15}) with IRAM-30m observations, we have also extracted spectra from regions that correspond to the IRAM-30m beam around the sources mm1 and mm3, as defined in Table \ref{coordinates} (considering mm1a and mm1b at the same coordinates). This was also done in order to show how much the $^{14}$N/$^{15}$N ratios change going to smaller scales with respect to those observed with single-dish observations. The IRAM-30m beam at 93 GHz is $\sim$27.6$\arcsec$. In order to be consistent with Fontani et al.~(\citeyear{fontani15}), we have fitted our spectra fixing the $T_{\rm ex}$ equal to those used in their work: 43 K for mm1 and 18 K for mm3. We noted that even though the two main beams are partially overlapping, the dominant emission of each of the two  is from the corresponding central source. The extracted spectra from regions equivalent to the IRAM-30m beam are shown in Fig.~\ref{spectra-30m} and the results from the fit are listed in Table \ref{30m-fit}.

\begin{table}
\begin{center}
\caption{$^{14}$N/$^{15}$N ratios derived from \15N towards P1a, P2a, P3a and P4a (left panel) and from \N15 towards P1b, P2b, P3b and P4b (right panel).}
\begin{tabular}{lcclcc}
\tabularnewline \hline \hline 
Source &$T_{\rm ex}$ &$\frac{{\rm N}_{2}{\rm H}^{+}}{^{15}{\rm NNH}^{+}}$ & Source &$T_{\rm ex}$ & $\frac{{\rm N}_{2}{\rm H}^{+}}{{\rm N}^{15}{\rm NH}^{+}}$ \\
 & (K) &  &  & (K) &   \\
 \hline
P1a &  20 & 200$\pm$34  & P1b & 20 & 200$\pm$25  \\ 
      &  30 & 186$\pm$31     &  & 30 & 217$\pm$29  \\
      &  40 & 178$\pm$29      &  & 40 & 212$\pm$27  \\ 
      &  50 & 173$\pm$27      &  & 50 & 222$\pm$30  \\  
\hline
P2a & 20 & 180$\pm$18 & P2b & 20 & 129$\pm$17  \\ 
     & 30 & 171$\pm$17 &  & 30 & 120$\pm$15  \\
    & 40 & 156$\pm$16 &  & 40 & 108$\pm$13  \\ 
    & 50 & 154$\pm$15 &  & 50 & 113$\pm$14  \\  
\hline
P3a & 20 & 120$\pm$18 & P3b & 20 & 100$\pm$15  \\ 
      & 30 & 114$\pm$17 &  & 30 & 100$\pm$16  \\
      & 40 & 111$\pm$16 &  & 40 & 109$\pm$17  \\ 
      & 50 & 109$\pm$15 &  & 50 & 100$\pm$15  \\  
\hline
P4a & 20 & 117$\pm$14 & P4b & 20 & 187$\pm$33  \\ 
      & 30 & 112$\pm$14 &  & 30 & 182$\pm$31  \\
      & 40 & 120$\pm$15 &  & 40 & 185$\pm$33  \\ 
      & 50 & 108$\pm$12 &  & 50 & 185$\pm$32  \\  
\bottomrule
\normalsize
  \label{ratio-table1}
  \end{tabular}
  \end{center}
  \end{table}

\subsection{$^{14}$N/$^{15}$N ratios}
We have computed the $^{14}$N/$^{15}$N ratios for \NN, along with the uncertainties derived propagating the error on total column densities. In Tables \ref{ratio-table1}, \ref{ratio-table2} and \ref{ratio-table3} the $^{14}$N/$^{15}$N ratios for the different regions defined in Sect.~\ref{molecule-map} are given. Since the ratios are obtained from column density values obtained from observations at different spectral resolution, we have carefully evaluated the effects of this data inhomogeneity in Appendix \ref{test}. There we show that the different spectral resolution does not affect significantly the column densities, and hence the $^{14}$N/$^{15}$N ratios.


In general, we have found values that span from $\sim$100 up to $\sim$220, where the higher values are comparable with those observed with the IRAM-30m (Fontani et al.~\citeyear{fontani15}). Moreover, the assumption of different $T_{\rm ex}$ does not change the results, that are consistent within the errors. Therefore, in the next session we will discuss the final results considering a single excitation temperature. Moreover, a different $T_{\rm ex}$ can change the column densities of each molecule, but does not influence their ratios, since the dependence on $T_{\rm ex}$ in the ratio is irrelevant when the two column densities are both computed from the (1--0) transition  and the lines are not optically thick. This is supported by the fact that the ratios given in Table \ref{ratio-table1} are the same within the uncertainties, towards each region.

\begin{table}
\begin{center}
\caption{$^{14}$N/$^{15}$N ratios derived from \15N and \N15 towards I1, I2, I3 and I4.}
\begin{tabular}{lccclccc}
\tabularnewline \hline \hline
 Source &$T_{\rm ex}$ &\multicolumn{2}{c}{$\frac{{\rm N}_{2}{\rm H}^{+}}{^{15}{\rm NNH}^{+}}$} & \multicolumn{2}{c}{$\frac{{\rm N}_{2}{\rm H}^{+}}{{\rm N}^{15}{\rm NH}^{+}}$}  \\
  & (K) & &   \\
 \hline
I1 & 20 & \multicolumn{2}{c}{220$\pm$32} & \multicolumn{2}{c}{275$\pm$50}  \\ 
   & 30 & \multicolumn{2}{c}{214$\pm$30} & \multicolumn{2}{c}{250$\pm$41}  \\ 
   & 40 & \multicolumn{2}{c}{200$\pm$27} & \multicolumn{2}{c}{257$\pm$46}  \\ 
   & 50 & \multicolumn{2}{c}{200$\pm$28} & \multicolumn{2}{c}{244$\pm$40}  \\ 
    \hline
I2 & 20 & \multicolumn{2}{c}{167$\pm$17} & \multicolumn{2}{c}{143$\pm$18}  \\ 
   & 30 & \multicolumn{2}{c}{150$\pm$15} & \multicolumn{2}{c}{120$\pm$15}  \\ 
   & 40 & \multicolumn{2}{c}{136$\pm$13} & \multicolumn{2}{c}{125$\pm$16}  \\ 
   & 50 & \multicolumn{2}{c}{146$\pm$14} & \multicolumn{2}{c}{127$\pm$16}  \\ 
    \hline
I3 & 20 & \multicolumn{2}{c}{100$\pm$13} & \multicolumn{2}{c}{86$\pm$12}  \\ 
   & 30 & \multicolumn{2}{c}{114$\pm$16} & \multicolumn{2}{c}{89$\pm$13}  \\ 
   & 40 & \multicolumn{2}{c}{111$\pm$15} & \multicolumn{2}{c}{91$\pm$14}  \\ 
   & 50 & \multicolumn{2}{c}{109$\pm$15} & \multicolumn{2}{c}{86$\pm$12}  \\ 
    \hline
I4 & 20 & \multicolumn{2}{c}{100$\pm$14} & \multicolumn{2}{c}{140$\pm$24}  \\
   & 30 & \multicolumn{2}{c}{100$\pm$14} & \multicolumn{2}{c}{150$\pm$29}  \\
   & 40 & \multicolumn{2}{c}{100$\pm$15} & \multicolumn{2}{c}{137$\pm$25}  \\
   & 50 & \multicolumn{2}{c}{93$\pm$14}  &  \multicolumn{2}{c}{130$\pm$23} \\
\bottomrule
\normalsize
  \label{ratio-table2}
  \end{tabular}
  \end{center}
  \end{table}

\begin{table}
\begin{center}
\caption{$^{14}$N/$^{15}$N ratios derived from \15N and \N15 for D1, D2 and D3 (upper panel). In the lower panel the  $^{14}$N/$^{15}$N ratios derived from both \15N and \N15 towards mm1 and mm3 in regions equivalent to the IRAM-30m beam are shown.}
\begin{tabular}{l*{3}{c}}
\tabularnewline \hline \hline    
Source &$T_{\rm ex}$ &$\frac{{\rm N}_{2}{\rm H}^{+}}{^{15}{\rm NNH}^{+}}$ & $\frac{{\rm N}_{2}{\rm H}^{+}}{{\rm N}^{15}{\rm NH}^{+}}$  \\
  & (K) & &   \\
 \hline
D1 & 20 & $\geq$245 & $\geq$204  \\ 
   & 30 & $\geq$231 & $\geq$188  \\ 
   & 40 & $\geq$242 & $\geq$200  \\ 
   & 50 & $\geq$250 & $\geq$204  \\ 
   \hline
D2 & 20 & 336$\pm$96 & $\geq$154  \\ 
   & 30 & 327$\pm$91 & $\geq$148  \\ 
   & 40 & 316$\pm$87 & $\geq$143  \\ 
   & 50 & 292$\pm$77 & $\geq$140  \\ 
   \hline
D3 & 20 & $\geq$243 & $\geq$340  \\ 
   & 30 & $\geq$250 & $\geq$321 \\ 
   & 40 & $\geq$261 & $\geq$353 \\ 
   & 50 & $\geq$250 & $\geq$333 \\ 
    \hline
     \hline  
     Source &$T_{\rm ex}$ &$\frac{{\rm N}_{2}{\rm H}^{+}}{^{15}{\rm NNH}^{+}}$ & $\frac{{\rm N}_{2}{\rm H}^{+}}{{\rm N}^{15}{\rm NH}^{+}}$  \\
  & (K) & &   \\
  \hline
mm1 & 43 & 219$\pm$22 &  180$\pm$33 \\ 
mm3 & 18 & 217$\pm$23 &  192$\pm$32 \\ 
\bottomrule
\normalsize
  \label{ratio-table3}
  \end{tabular}
  \end{center}
  \end{table}

\section{Discussion}
\label{discussion}

\begin{figure*}
\centering
\includegraphics[width=43pc]{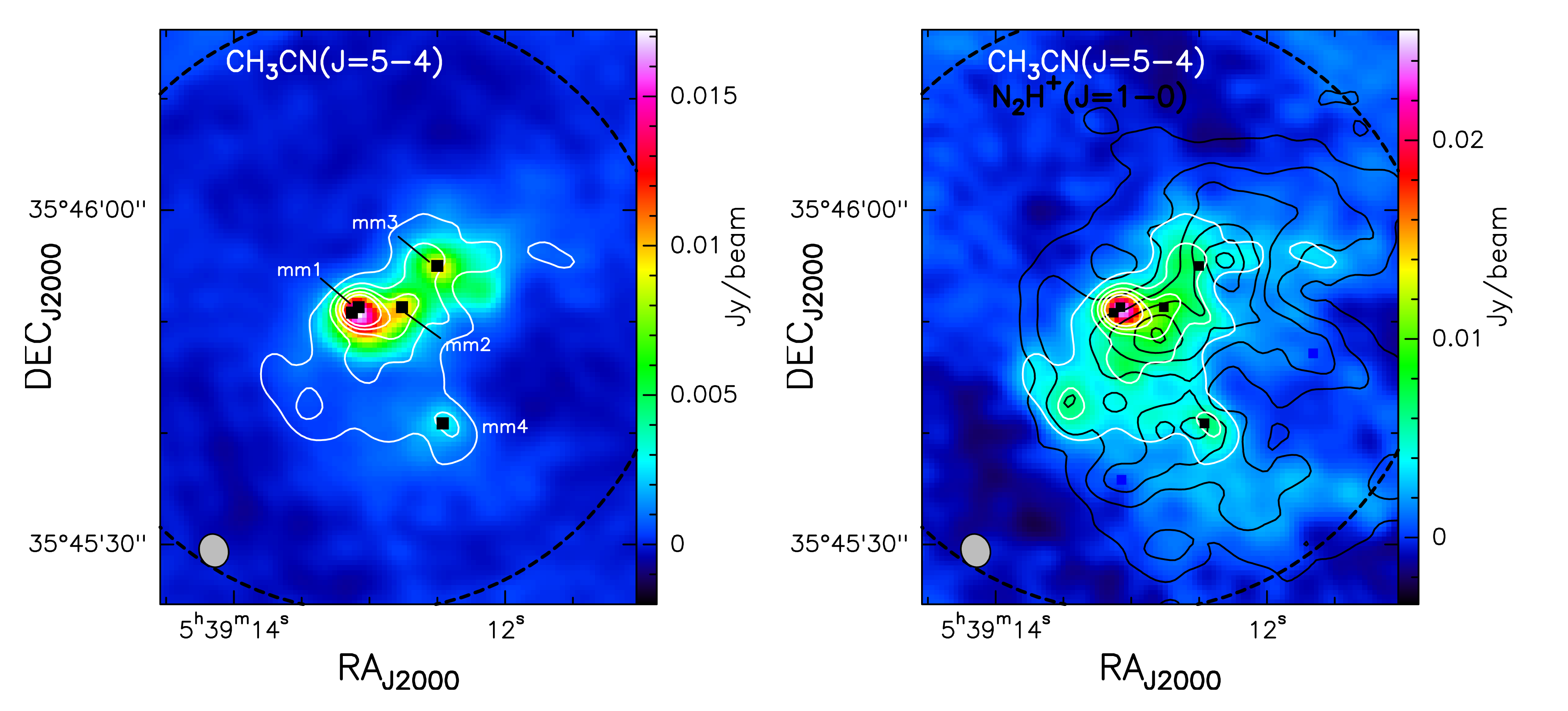}
\caption{\emph{Left panel}: The colours are the 3mm-continuum intensity (the same of Fig.~\ref{cont-map}). The white contour levels are 5, 10, 15, 20 and 25 times the 1$\sigma$ rms of the CH$_{3}$CN(5--4) map, equal to $\sim$0.6 mJy/beam km s$^{-1}$. \emph{Right panel}: The colours are the averaged emission of CH$_{3}$CN(5--4), the white contours are the same as in the left panel and the black contours are the same as in Fig.~\ref{n2hp-map}. In both panels the black squares indicate the positions of the continuum sources as in Fig.~\ref{n2hp-map} (black). The dashed circle represents the NOEMA field of view and the synthesized beam is the ellipse indicated in the lower left corner.}
\centering
\label{ch3cn-map}
\end{figure*}

\subsection{Comparison between line and continuum emission maps}
\label{discussion-map}
The averaged emission maps of the J=1--0 transition of \NN, \15N and \N15 (Fig.~\ref{n2hp-map} and \ref{15n-map}) present a shift of $\sim$2-3$\arcsec$ to the south-west with respect to the continuum sources. In particular this effect is stronger for mm1, mm2 and mm3 than for mm4. The overall structure of the integrated \NN(1--0) (and its $^{15}$N-isotopologues) consists of three main cores. Two of these are located to the south of the mm1 and mm2 continuum sources, and to the west of the mm3 continuum sources, while the third is located exactly to the mm4 source. Numerous studies have shown that during the collapse of a low-mass core, the CO desorbs from the grain mantles, because of the rising of the temperature, and this causes a substantial destruction of \NN (e.g. Di Francesco et al.~\citeyear{difrancesco04}). The chemical processes at work in high-mass star-forming regions are likely different from those in low-mass star-forming ones, but the destruction of \NN\:by CO should occur in both cases. Towards mm1 the presence of a young massive protostar is well-known (see Sect.~\ref{observations}) and the displacement between the dust and \NN peak suggests that the heating of the protostar, and probably of the molecular outflows associated with it (Beuther et al.~\citeyear{beuther02a}), may have caused the desorption of CO from grain mantles and the subsequent destruction of \NN. Probably something similar is happening also around mm3 although we do not have clear evidences that star formation already started. Moreover, the correspondence of the dust and \NN peak towards mm4 is an evidence that it is chemically less evolved and hence maybe starless. Something similar for high-mass star-forming regions has been reported towards IRAS 23033+5951 (Reid \& Matthews \citeyear{reid08}) and towards the massive protocluster AFGL 5142 (Busquet et al.~\citeyear{busquet11}). To investigate better this situation we have searched in our spectral setup for other molecules which could better trace the continuum sources and the associated star formation activity. In particular, in Fig.~\ref{ch3cn-map} the averaged emission map of CH$_{3}$CN(J=5--4, K=0,1) is shown. In the left panel the clear correspondence between the 3mm-continuum sources and CH$_{3}$CN can be noted; moreover, in the right panel it is evident that CH$_{3}$CN and \NN do not coincide towards mm1, mm2 and mm3, while the core-structure defined by the two different molecules corresponds towards mm4.
Moreover, the asymmetry of \NN around the millimeter continuum sources could be due to the presence of the interaction with the gas of the complex group of \HII regions located in the North-East. However, we do not have observational proof of this point and it needs to be tested.

As already mentioned in Sect.~\ref{molecule-map}, in the western and in the south-eastern part of the cluster, two other \NN-emission zones are present. In particular, A corresponds to a core structure in the $^{15}$N-isotopologues maps (P4a and P4b in Fig.~\ref{15n-map}) and corresponds also to one of the emission peaks of the high-density tracer H$^{13}$CO$^{+}$ defined in Fig.~8 of Beuther et al.~(\citeyear{beuther02a}), which is a high-density tracer. However, B presents a more complicated structure, probably related with the presence of multiple outflows in the whole region (outflow \emph{A} in Beuther et al.~\citeyear{beuther02a}). Leurini et al.~(\citeyear{leurini11}) discussed the \NN(1--0) emission towards the IRDC G351.77-0.51 and they found a velocity gradient on large scale, probably associated with outflows detected in CO. Although \NN is seldom associated with outflows, this could be due to an in interaction of outflow shocks with the molecular envelope (traced also by \NN). A similar situation was detected in the same transitions towards the class 0 object IRAM 04191+1522 by Lee et al.~(\citeyear{lee05}). Moreover, Codella et al.~(\citeyear{codella13}) present the first detection of \NN towards a low-mass protostellar outflow, L1157-B1, a bow-shock, at $\sim$0.1 pc from the protostellar cocoon L1157.

\subsection{Is N-fractionation a core-scale effect?}
\label{discussion-corescale}
\begin{figure*}
\centering
\includegraphics[width=43pc]{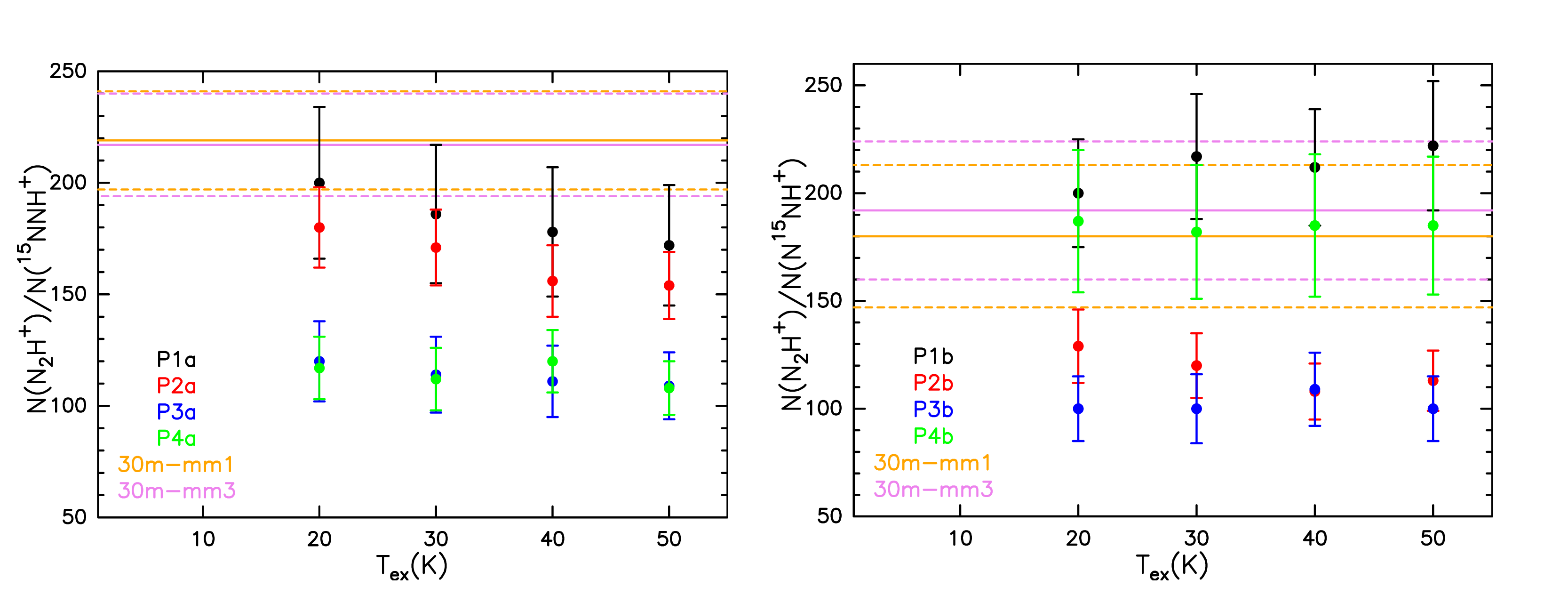}
\caption{\emph{Left panel}: $^{14}$N/$^{15}$N ratios obtained from \15N, as a function of the different $T_{\rm ex}$ assumed, towards P1a (black points), P2a (red points), P3a (blue points) and P4a (green points). \emph{Right panel}: The same as the left panel but for \N15 towards P1b (black points), P2b (red points), P3b (blue points) and P4b (green points). In both panels the orange solid line is the value obtained in a polygon equal to the IRAM-30m beam towards mm1 (the dashed lines correspond to the error bar) and the pink dashed line is the same but towards mm3, for \15N in the left panel and \N15 in the right panel. The $T_{\rm ex}$ assumed to compute the $^{14}$N/$^{15}$N ratios toward 30m-mm1 and 30m-mm3 are equal to those derived in Fontani et al.~(\citeyear{fontani15}): 43 K for mm1 and 18 K for mm3.}
\centering
\label{15n-30m}
\end{figure*}

Fontani et al.~(\citeyear{fontani15}) have derived, with the IRAM-30m radiotelescope, $^{14}$N/$^{15}$N ratios of 190$\pm$20 with \15N and of 180$\pm$23 with \N15 towards mm1, and of 210$\pm$12 with \15N and of 180$\pm$13 with \N15 towards mm3. Extracting the spectra in regions equivalent to the IRAM-30m beam at 93 GHz, we have derived the $^{14}$N/$^{15}$N ratios listed in the lower panel of Table \ref{ratio-table3}. These are consistent with the results obtained with the single-dish telescope, within the errors. The comparison of these results with the $^{14}$N/$^{15}$N ratios obtained towards P1a, P2a, P3a and P4a and towards P1b, P2b, P3b and P4b is shown in Fig.~\ref{15n-30m}. In particular, it is evident from left panel of Fig.~\ref{15n-30m} that there is a $^{15}$N-enrichment (i.e. a $^{14}$N/$^{15}$N ratio significantly lower) towards P2a, P3a and P4a. The same happens towards P2b and P3b (right panel of Fig.~\ref{15n-30m}). This means that at higher angular resolution, where the core scales can be resolved, it is possible to reveal a different N-fractionation with respect to that observed with a single-dish telescope and more representative of the average $^{14}$N/$^{15}$N between the cores and the envelope of the whole star-forming region. This suggests that the mechanisms that produce more $^{15}$N with respect to $^{14}$N in \NN can occur in smaller regions and could be a local effect. The fact that towards P1a and P1b, $^{14}$N/$^{15}$N values consistent with IRAM-30m observations ($\sim$200) are found could be explained by the fact that these regions are near an evolved high-mass source (mm1) and the chemistry could be different toward hotter gas. Moreover, the different $^{14}$N/$^{15}$N ratios between P4a and P4b, $\sim$115 and $\sim$185, respectively, could be due to a different behaviour of the two $^{15}$N-isotopologues, but we will discuss about this in the next section.

In general, the lower $^{14}$N/$^{15}$N ratios of $\sim$100--110, are similar to the values measured by Guzmán et al.~(\citeyear{guzman17}) toward protoplanetary discs in HCN, and they are also consistent with the low values measured in pristine Solar System materials. However, N-fractionation of different molecules, like HCN and \NN, could be different. To investigate better this point, consistent measurements of
N-fractionation from different species in the same sources are
needed to understand the puzzle of nitrogen isotopic ratios.
Moreover, high-mass angular resolution observations towards other high-mass star-forming regions are needed in order to confirm these results. In fact, it is also important to gather more data in sources that are good candidates to represent the environment in which our Sun was born. In this respect, intermediate- and high-mass star-forming cores are interesting targets because the Sun was probably born in a rich cluster that also contained massive stars (e.g. Adams \citeyear{adams10}, Lichtenberg et al.~\citeyear{lichtenberg19}).

Unfortunately, none of the published chemical models up to now are able to reproduce the low $^{14}$N/$^{15}$N observed in this work. Moreover, they are not able to reproduce observed values of N-fractionation in \NN in general. However, they are gas-phase models that feature only low-temperature exchange reactions as the only fractionation mechanism (e.g. Roueff et al.~\citeyear{roueff15}, Wirström \& Charnley \citeyear{wirstrom18}). Probably other mechanisms must be taken into account, like the isotope selective photodissociation of N$_{2}$ and grain surface chemistry (Furuya \& Aikawa et al.~\citeyear{furuya18}). In particular, these authors have proposed some mechanisms to explain the $^{15}$N-depletion in \NN in prestellar cores starting from their parental clouds. They found that during the evolution of a molecular cloud, where the external UV radiation field is not fully shielded, the $^{14}$N$^{15}$N molecule could be photodissociated where the $^{14}$N$_{2}$ is not, causing an enrichment of atomic $^{15}$N. The subsequent adsorption of atomic N creates an enrichment of $^{15}$N in NH$_{3}$ ice by surface reactions. They found that as long as the photodesorption of NH$_{3}$ ice is less efficient than that of N$_{2}$ ice, the net effect is the loss of $^{15}$N from the gas phase. However, these results are based on a specific physical model adapted for cloud with density of $\sim$10$^{4}$cm$^{-3}$, which could not be appropriate for dense sources. To understand the existing $^{15}$N-observations more numerical studies of $^{15}$N-fractionation are needed.


\subsection{\15N vs \N15}
In this section we will discuss about the possible differences between the two $^{15}$N-isotopologues of \NN: \15N and \N15. In order to do that, we have compared the $^{14}$N/$^{15}$N ratios obtained from \15N and \N15 towards I1, I2, I3 and I4, since the column densities of the two $^{15}$N-isotopologues were derived from the same regions.

\begin{figure}
\centering
\includegraphics[width=20pc]{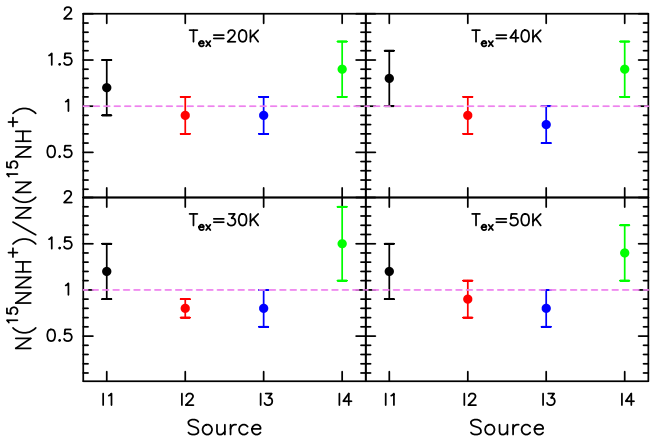}
\caption{\15N/\N15 ratios towards I1 (black points), I2 (red points), I3 (blue points) and I4 (green points). Different panels show the results with the different $T_{\rm ex}$ assumed. The pink solid line represents the locus of points where \15N/\N15=1.}
\centering
\label{15n-n15}
\end{figure}

In Fig.~\ref{15n-n15} the \15N/\N15 column density ratios, for the different excitation temperatures assumed, are shown. In this case, the dependence on the $T_{\rm ex}$ is small because both the $^{15}$N-isotopologues are almost optically thin. We have found that \15N is more abundant (at 1$\sigma$ level) with respect to \N15 towards I4, and the contrary towards I1, I2 and I3.
Concerning what happens towards I4, it is difficult to interpret our results from an evolutionary point of view, because the nature of this source is unknown (as already mentioned in Sect.~\ref{discussion-map}). In this case, the ratio between the two $^{15}$N-isotopologues is higher than one, also including the errorbar. This could be a clue that in some particular cases of densities and temperatures the two $^{15}$N-isotopologues could form through different reaction pathways. This finding should also be confirmed by a higher statistics.

As already mentioned, the most recent chemical models are not able to reproduce a different behaviour between the two $^{15}$N-isotopologues. Other works toward very different environments show a tentative trend of \15N to be less abundant than \N15: Kahane et al.~(\citeyear{kahane18}) towards OMC-2 FIR4 (\15N/\N15$\sim$0.8) and Redaelli et al.~(\citeyear{redaelli18}) towards the prestellar cores L1544 and L694-2 (\15N/\N15$\sim$0.9 and $\sim$0.8, respectively).

\subsection{$^{14}$N/$^{15}$N ratios in diffuse regions}
In Tab.~\ref{ratio-table3} the results obtained towards the more diffuse \NN emission regions D1, D2 and D3 are shown. All of these results are lower limits since the $^{15}$N-isotopologues were not detected, except for \15N towards D2. The lower limit values obtained are, on average, higher than 200--250. These are also the maximum values obtained towards the peak-emission regions. Only the lower limits towards D2 from \N15 are smaller, but given that the detection from \15N confirm a $^{14}$N/$^{15}$N ratio of $\sim$320, and that we have found a maximum difference of $\sim$1.5 (see previous Section), we can confirm that this lower limits corresponds to $^{14}$N/$^{15}$N ratios higher than 200.
These results mean that $^{15}$N tend to be less enriched in \NN in the outer parts of a star-forming region. This could be a possible confirmation of the work of Furuya et al.~(\citeyear{furuya18}): in NI/N$_{2}$ transition zones, where the gas is less dense and the interstellar UV radiation field is not significantly attenuated ($A_{\rm v}$<3 mag), the $^{15}$N is frozen in NH$_{3}$ ice and there is a loss of $^{15}$N from the gas phase.

\section{Conclusion}
\label{conclusions}
We used the IRAM NOEMA Interferometer to observe the emission of \NN(1--0) and its $^{15}$N-isotopologues towards the high-mass star-forming protocluster IRAS 05358+3543. The \NN dense gas emission consists of three main cores, two of which are displaced with respect to the 3mm continuum sources, probably because of the star-formation activity that causes the \NN destruction by CO desorption from grain mantles. We have also discussed the presence of two other \NN-emission zones: A and B. The first define a fourth core from the $^{15}$N-isotopologues emission maps and corresponds to a H$^{13}$CO$^{+}$ peak observed in previous works. The second is probably associated with an outflow coming out from the more evolved region in the north.

The $^{14}$N/$^{15}$N ratios derived toward regions that correspond to the IRAM-30m beam are higher than those from the cores of the cluster. This indicates the possibility that $^{15}$N-enrichment in star-forming regions is a local effect. The lower values of $\sim$100 toward the cores are similar to some values measured in protoplanetary discs, and also to the low values measured in the pristine Solar System material. Moreover, the $^{14}$N/$^{15}$N ratios derived from \15N and \N15 show some differences. In particular, towards I1 and I4 \15N is more abundant than \N15. However, the most recent chemical models are not able to reproduce the low $^{14}$N/$^{15}$N values found for \NN in this work and in other works.

Finally, the $^{14}$N/$^{15}$N upper limits derived in the more diffuse \NN emission regions of the cluster, point to values higher than those derived in the cores. This confirms the conclusion that N-fractionation, measured from \NN, seems to be a local effect, at least in this high-mass star-forming region. However, more observations, also of different molecules, are needed to unveil which physical parameters are affecting the N-fractionation, thus helping us to gain understanding on the chemistry involved.

\section*{Acknowledgements}
This work is based on observations carried out under projects number S16AK and 081-16 with the IRAM NOEMA Interferometer and 30m telescope, respectively. IRAM is supported by INSU/CNRS (France), MPG (Germany) and IGN (Spain).
We are grateful to the IRAM NOEMA staff for their help during the calibration of the data and the IRAM-30m telescope staff for their help during the observations. 
Many thanks to the anonymous referee for the careful reading of the paper and the comments that improved the work.
LC acknowledges support from the Italian Ministero
dell’Istruzione, Università e Ricerca through the grant Progetti Premiali 2012 - iALMA (CUP C52I13000140001). PC acknowledges support from the European Research Council (project PALs 320620).
 LC sincerely acknowledge the hospitality of the Joint ALMA Observatory (JAO), where significant part of the manuscript was written.






{}

\twocolumn


\newpage
\appendix

\section{Extracted spectra}
\label{spectra}
In this appendix, the extracted spectra of the J=1--0 transition of \NN, \15N and \N15 from the regions P1a--P4a, P1b--P4b, I1, I2, I3, I4, D1, D2 and D3 are shown. The definition of the regions can be found in Sect.~\ref{results}.

\onecolumn
\begin{figure}
\centering
\includegraphics[width=44pc]{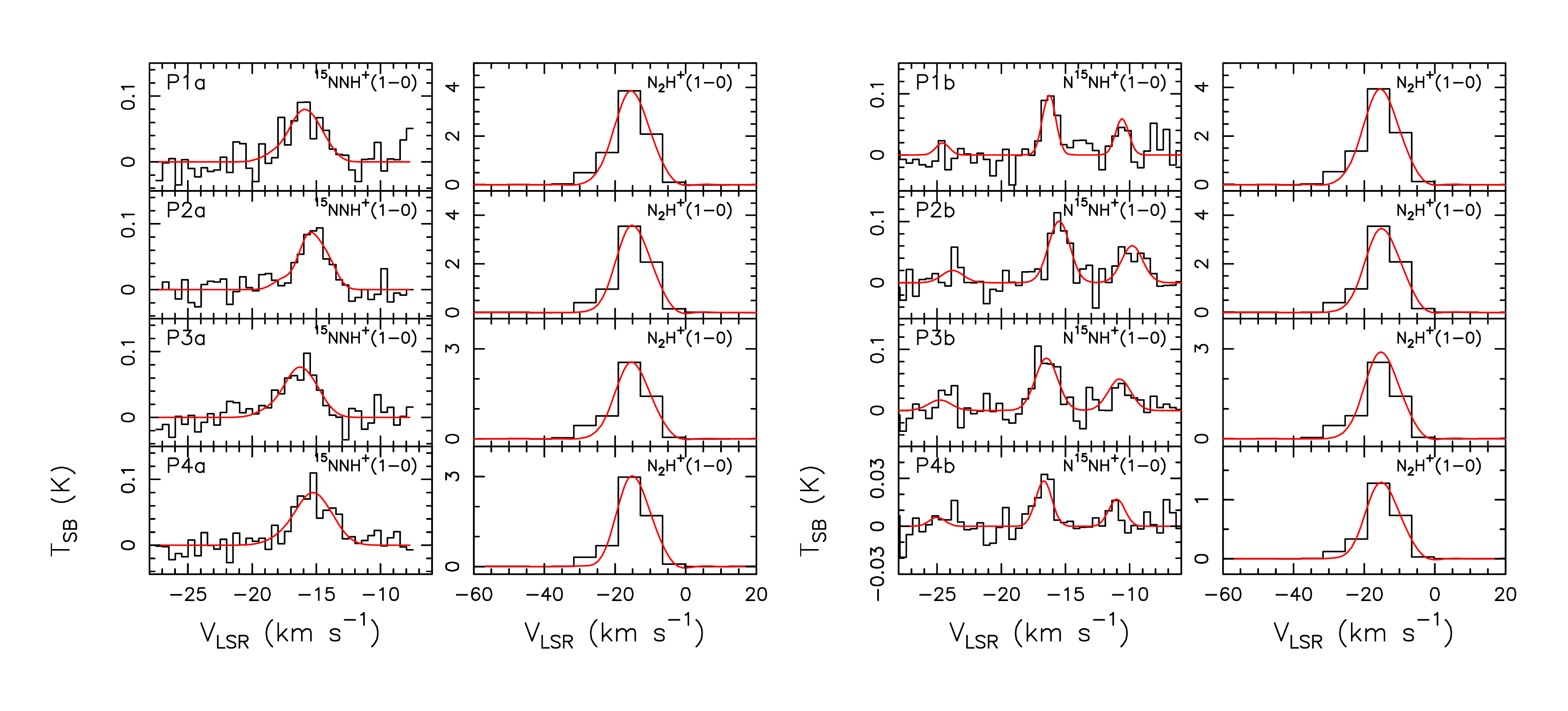}
\caption{\emph{Left panel}: Spectra of the \15N(1--0) and \NN(1-0) transitions (first and second column, respectively) obtained for the P1a, P2a, P3a and P4a regions (first, second, third and fourth line, respectively). \emph{Right panel}: Spectra of the \N15(1--0) and \NN(1-0) transitions (first and second column, respectively) obtained for the P1b, P2b, P3b and P4b regions (first, second, third and fourth line, respectively). 
For each spectrum, in both panels, the x-axis represents the systematic velocity of the source. The y-axis shows the intensity in synthesised beam temperature units. The red curves are the best Gaussian fits obtained with MADCUBA with a $T_{\rm ex}$=30K.}
\centering
\label{spectra-P}
\end{figure}

\begin{figure}
\centering
\includegraphics[width=40pc]{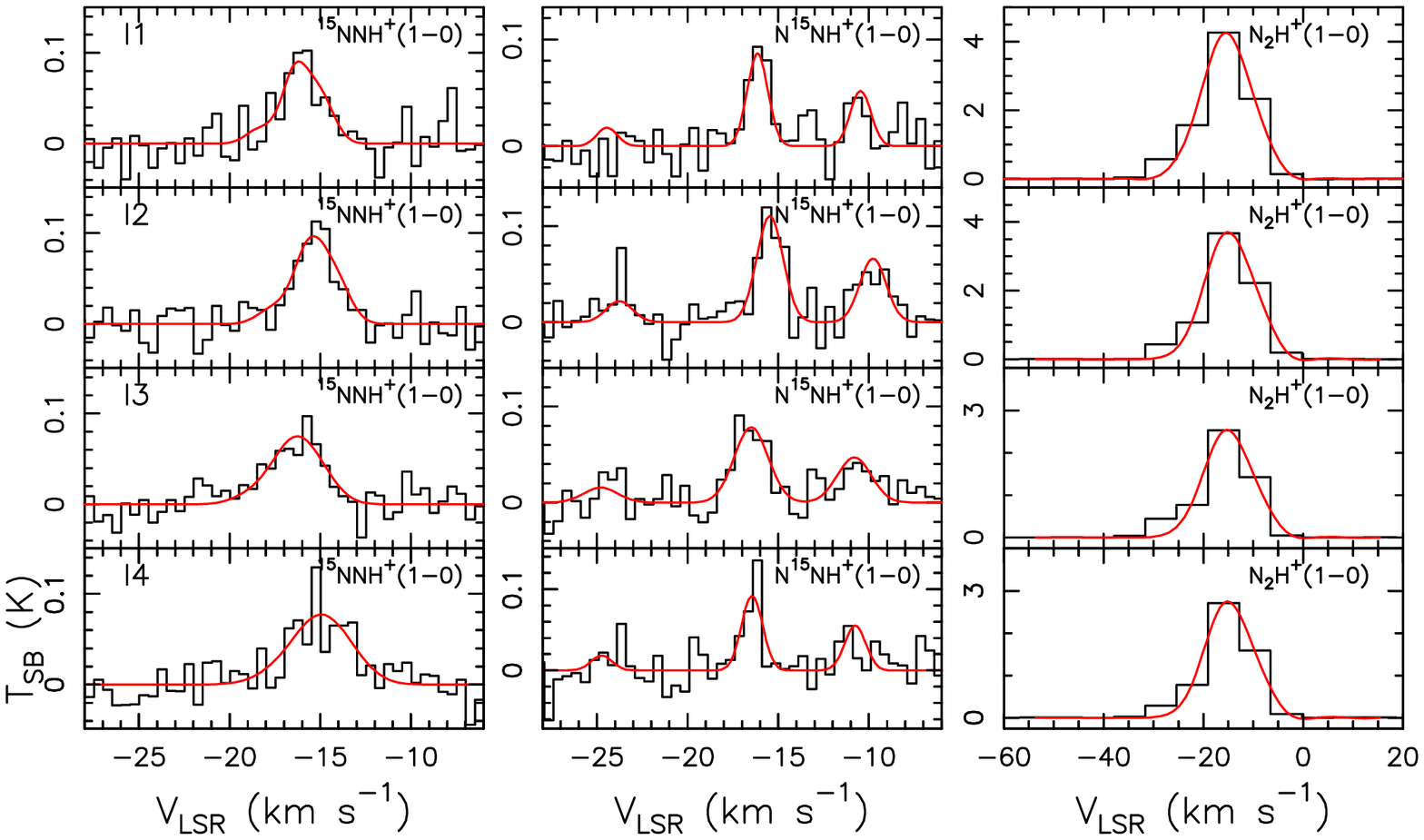}
\caption{Spectra of the \15N(1--0), \N15(1--0) and \NN(1-0) transitions (first, second and third column, respectively) obtained for the I1, I2, I3 and I4 regions (first, second, third and fourth line, respectively). For each spectrum the x-axis represents the systematic velocity of the source. The y-axis shows the intensity in synthesised beam temperature units. The red curves are the best Gaussian fits obtained with MADCUBA with a $T_{\rm ex}$=30K.}
\centering
\label{spectra-inters}
\end{figure}

\begin{figure}
\centering
\includegraphics[width=40pc]{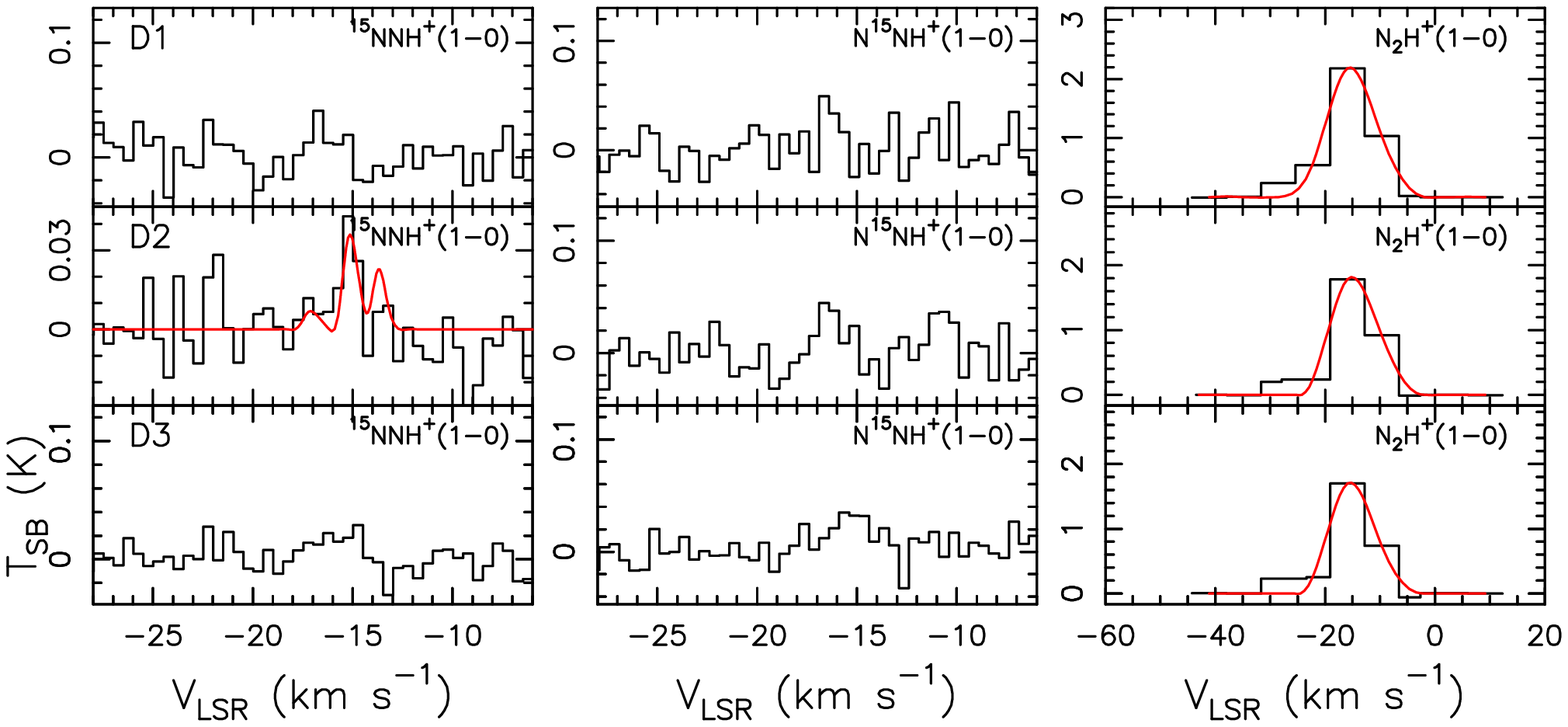}
\caption{Spectra of the \15N(1--0), \N15(1--0) and \NN(1-0) transitions (first, second and third column, respectively) obtained for the D1, D2 and D3 regions (first, second and third line, respectively). For each spectrum the x-axis represents the systematic velocity of the source. The y-axis shows the intensity in synthesised beam temperature units. The red curves are the best Gaussian fits obtained with MADCUBA with a $T_{\rm ex}$=30K.}
\centering
\label{spectra-diffuse}
\end{figure}

\begin{figure}
\centering
\includegraphics[width=40pc]{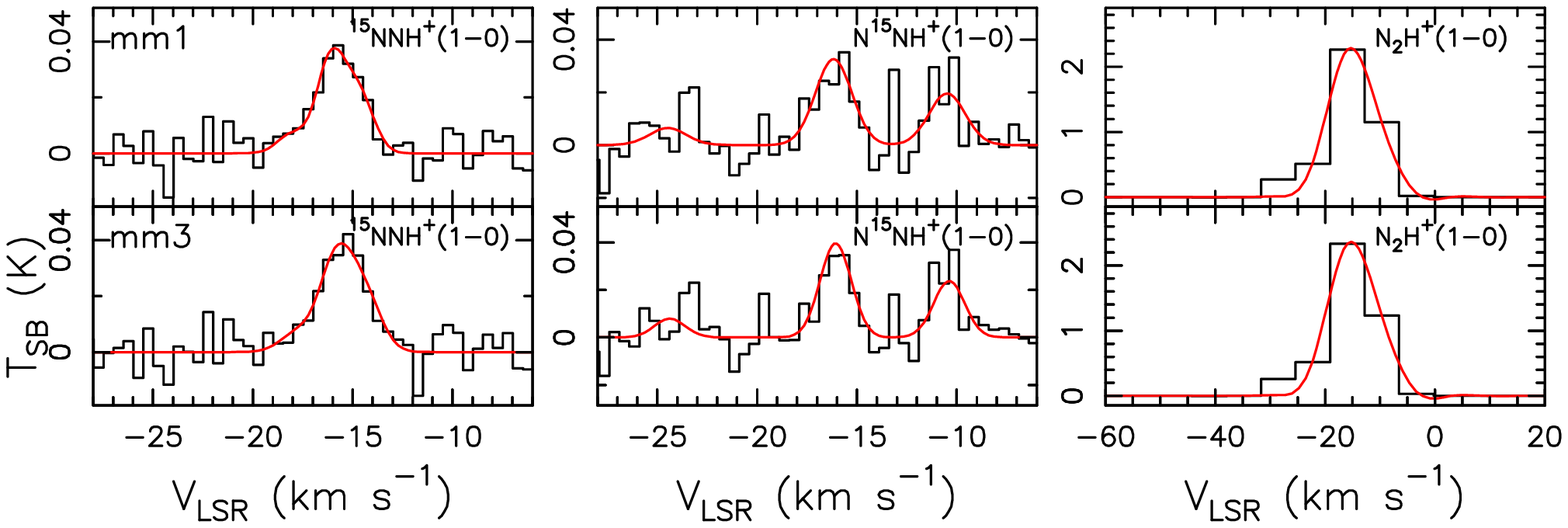}
\caption{Spectra of the \15N(1--0), \N15(1--0) and \NN(1-0) transitions (first, second and third column, respectively) obtained for the regions that correspond to the IRAM-30m beam around mm1 and mm3 (first and second line, respectively). For each spectrum the x-axis represents the systematic velocity of the source. The y-axis shows the intensity in synthesised beam temperature units. The red curves are the best Gaussian fits obtained with MADCUBA with a $T_{\rm ex}$=43 K for mm1 and a $T_{\rm ex}$=18 K for mm3, that are equal to those derived in Fontani et al.~(\citeyear{fontani15}).}
\centering
\label{spectra-30m}
\end{figure}

\newpage
\twocolumn
\section{Spectra simulation tests}
\label{test}
In this appendix we have estimated how much our results are influenced by the different spectral resolution between the main isotopologue (\NN, of about 6.45 km/s), and the $^{15}$N-isotopologues (\15N and \N15, of about 0.5 km/s). For this test we have taken into account the spectra extracted from P1a and P1b, and the corresponding fit results with a $T_{\rm ex}$=20 K

\subsection{$^{15}$N-isotopologues at lower spectral resolution}
\begin{figure}
\centering
\includegraphics[width=20pc]{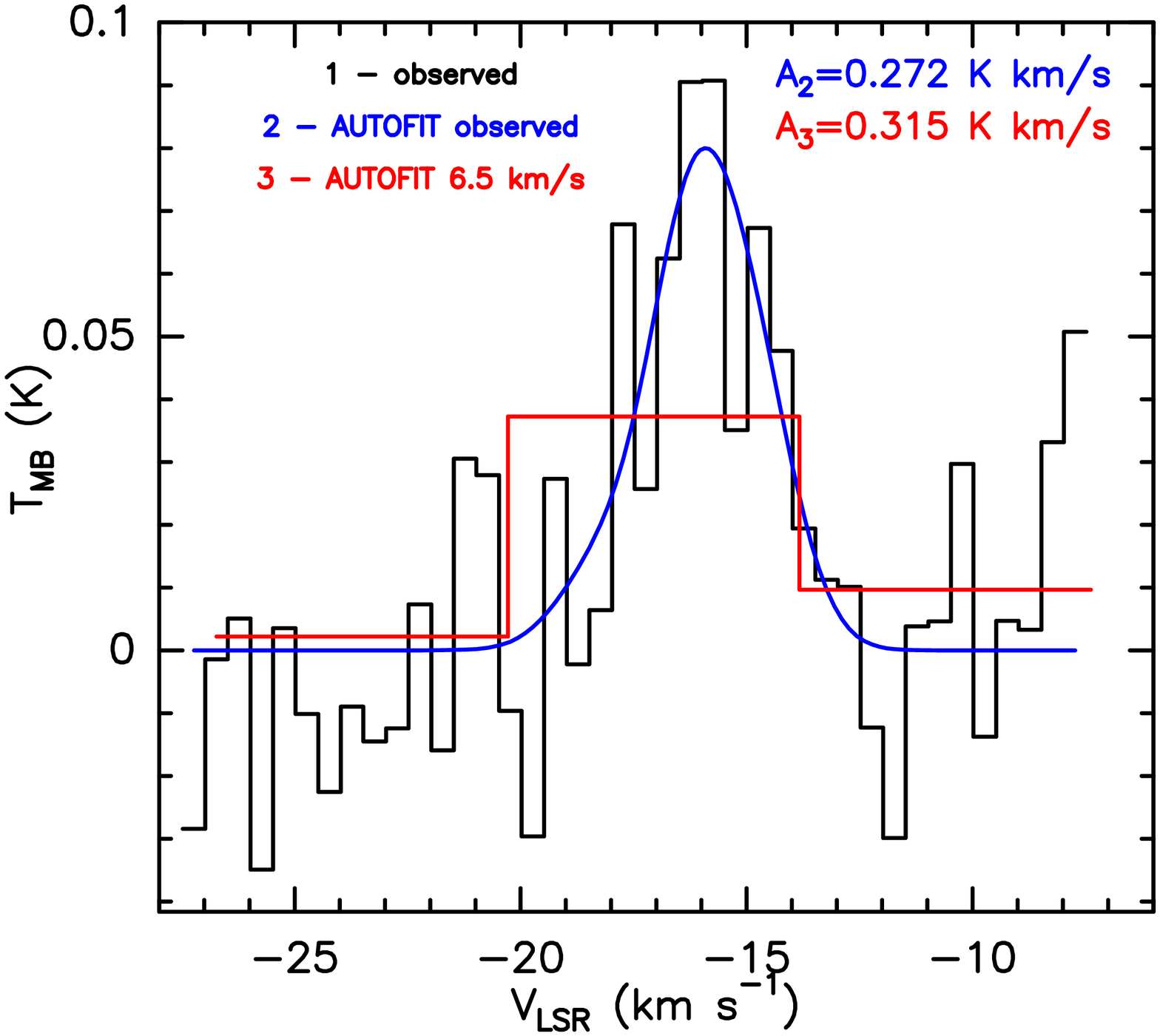}
\caption{\15N(1--0) spectra. The observed spectrum is the black one, the best fit to the observed spectrum is the blue solid line and the smoothed spectrum to 6.5 km/s is the red one.}
\centering
\label{15n-lowres}
\end{figure}
\begin{figure}
\centering
\includegraphics[width=20pc]{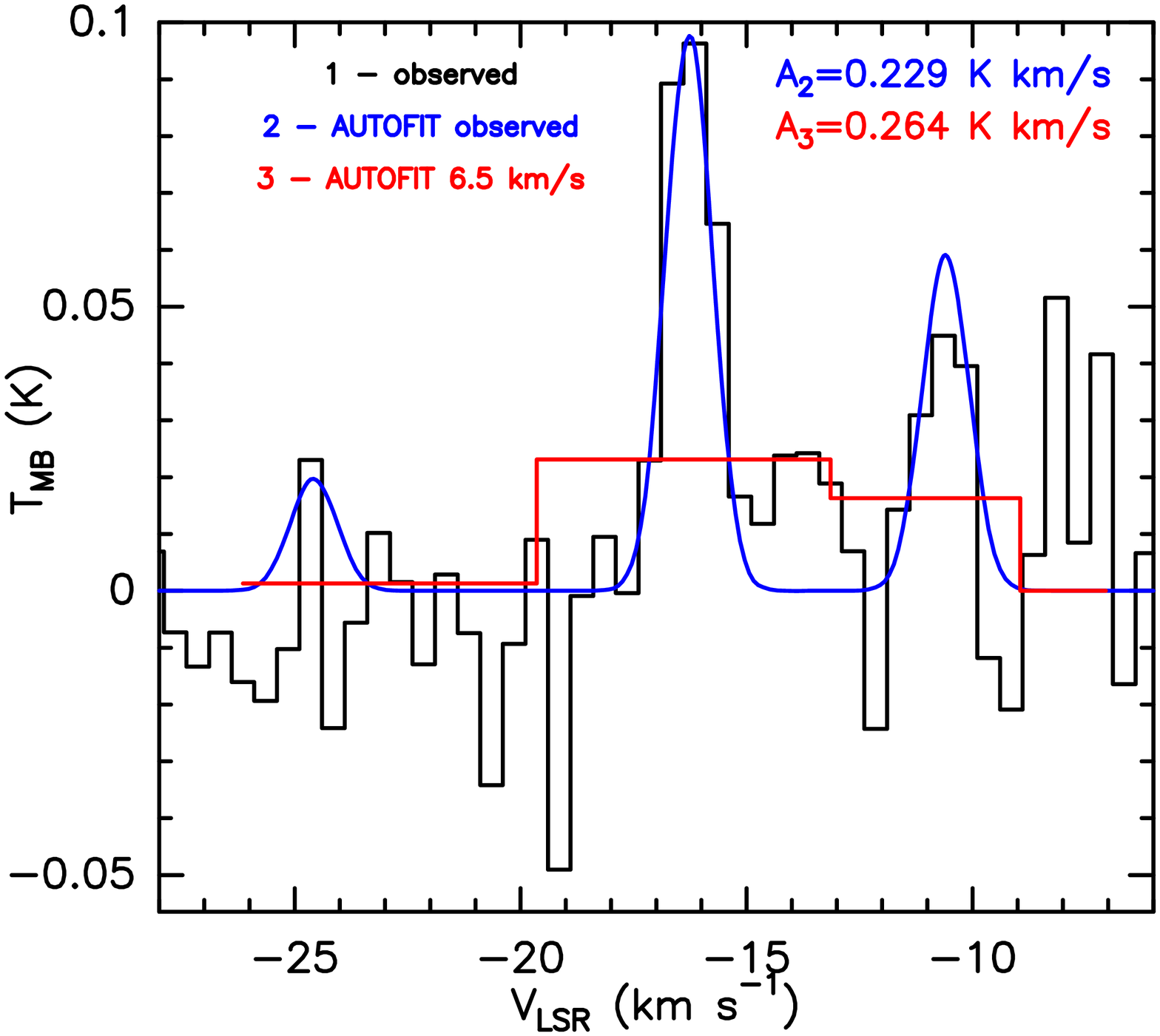}
\caption{\N15(1--0) spectra. The observed spectrum is the black one, the best fit to the observed spectrum is the blue solid line and the smoothed spectrum to 6.5 km/s is the red one.}
\centering
\label{n15-lowres}
\end{figure}

We have smoothed the \15N(1--0) transition to a 6.5 km/s resolution. In Fig.~\ref{15n-lowres}, the black spectrum is the observed one, the blue solid line is the fit to the observed spectrum, and the red line is the smoothed one. The integrated intensity ({\it A}=$\int T_{\rm MB}\, dv$) of the observed spectrum is (0.27$\pm$0.04) K km/s (relative error $\Delta A$/{\it A}$\simeq$15\%), while the integrated intensity of the smoothed spectrum is (0.31$\pm$0.06) K km/s, 13\% higher than the observed one. This error is comparable to the error given from the fit. We have done the same analysis with \N15(1--0) (Fig.~\ref{n15-lowres}), for which we have obtained, for the smoothed spectrum, an integrated intensity of (0.26$\pm$0.08) K km/s, 12\% higher than the observed one (0.23$\pm$0.03 K km/s), that again is comparable with the relative error given by the fit of the observed spectrum ($\Delta A$/{\it A}$\simeq$13\%). The test hence shows that the integrated intensity of the $^{15}$N-isotopologues, and thus their column densities, are not affected by changing the resolution from 0.5 to 6.5 km/s.

\subsection{\NN at higher spectral resolution}
\label{analysis-nn}
\begin{figure}
\centering
\includegraphics[width=20pc]{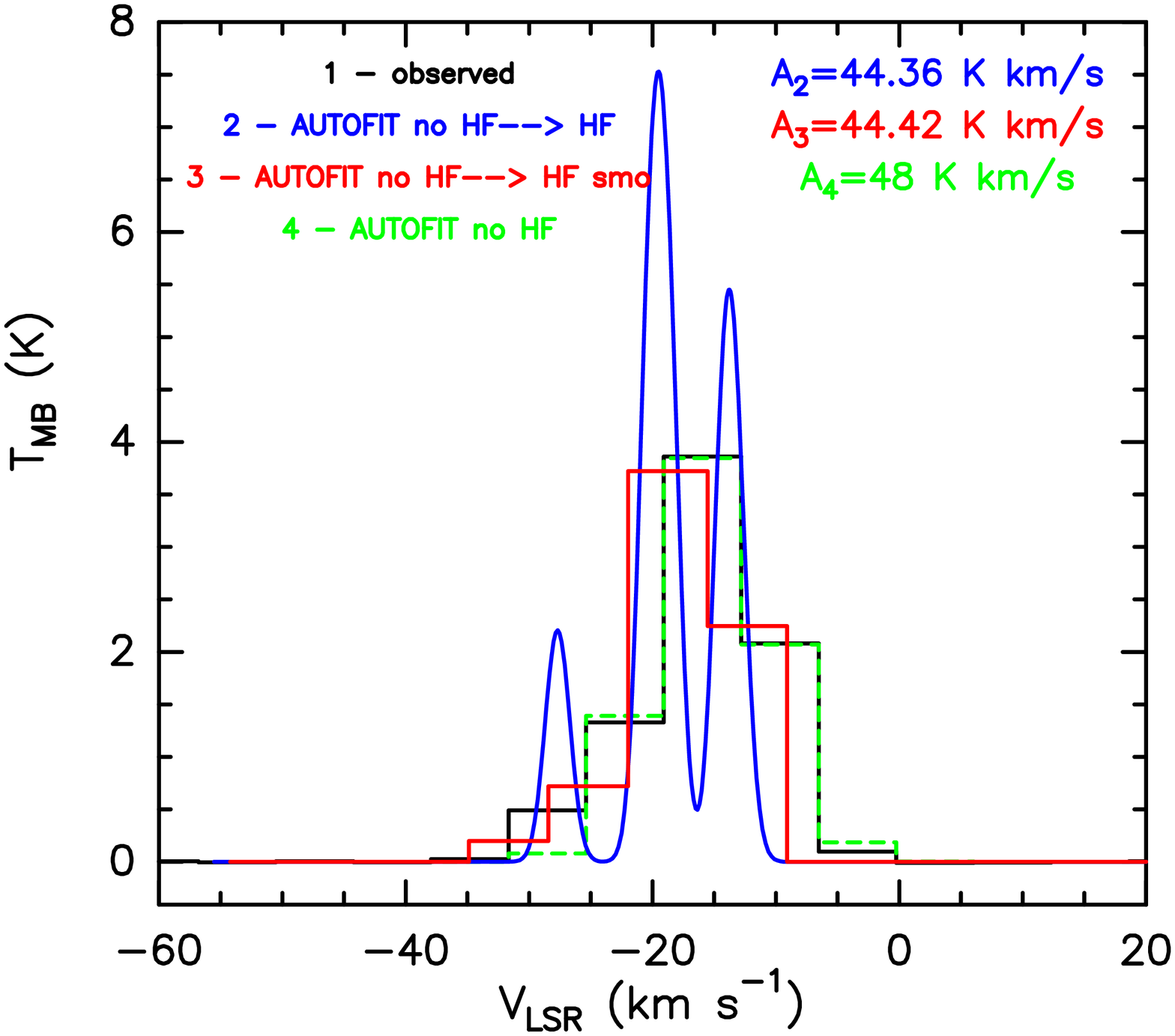}
\caption{\NN(1--0) spectra. The observed spectrum is the black one and the best fit to this is the green dashed line. The blue solid line is the simulated spectrum at high spectral resolution (0.5 km/s) with a line width of 2.3 km/s and the red one is the smooth of the simulation to 6.5 km/s.}
\centering
\label{nn-highres}
\end{figure}

We have simulated the spectrum of \NN(1--0) with a resolution of 0.5 km/s using the results of the fit obtained for a $T_{\rm ex}$=20 K, towards P1a, but with a line width of 2.3 km/s (which is the average value measured from \15N). The obtained spectrum is the blue in Fig.~\ref{nn-highres}, with an integrated intensity of (44$\pm$2) K km/s. Then, we have smoothed this simulated spectrum to 6.45 km/s obtaining the red one in the same figure. The integrated intensity of this latter is the same of that at higher spectral resolution. We can also note that the smoothed simulated spectrum is comparable with the observed one (black) that has an integrated intensity of (48$\pm$2) K km/s ($\Delta A$/{\it A}$\simeq$4\%), 8\% higher than the simulated one.
We can conclude that, even if we cannot resolve the hyperfine structure of \NN(1--0), with our analysis we have obtained an accurate integrated intensity, with at most 4\% overestimation.

\twocolumn
\section{Fit results}
\label{appendix-fit}
In this appendix, the results of the fitting procedure to the \NN(1--0), \15N(1--0) and \N15(1--0) transitions towards all the sources defined in Sect.~\ref{results}. The method is explained in Sect.~\ref{madcuba}.

\onecolumn
\begin{longtable}{l*{10}{c}}
\caption{\label{polygon-fit}Values obtained with the fitting procedure described in Sect.~\ref{madcuba} to the J=1--0 transition of \NN and \15N towards P1a, P2a, P3a and P4a (top panel of the table), and of the J=1--0 transition of \NN and \N15 towards P1b, P1b, P3b and P4b (bottom panel of the table).} 
\tabularnewline \hline \hline    
Source & \multicolumn{5}{c}{\NN(1--0)} & \multicolumn{5}{c}{\15N(1--0)} \\
  & $T_{\rm ex}$ & $FWHM^{a}$ &$v_{\rm LSR}$  & $\tau$  & $N_{\rm tot}$ & $T_{\rm ex}$ & $FWHM$ &$v_{\rm LSR}$  & $\tau^{b}$  & $N_{\rm tot}$  \\
 & (K) &(km s$^{-1}$) & (km s$^{-1}$) &  & ($\times$10$^{14}$cm$^{-2}$)  & (K) &(km s$^{-1}$) & (km s$^{-1}$) & & ($\times$10$^{11}$cm$^{-2}$)  \\
 \hline
  \endhead
 P1a & 20 & 11.1$\pm$0.6 & -15.2$\pm$0.3 & 0.26$\pm$0.02   & 1.0$\pm$0.1   & 20 &  2.2$\pm$0.7 &  -16.2$\pm$0.3   & 0.002$\pm$0.001     & 5$\pm$1   \\  
   & 30 & 11.3$\pm$0.6 & -15.2$\pm$0.3 & 0.16$\pm$0.02   & 1.3$\pm$0.2   & 30 &  2.2$\pm$0.7 &  -16.2$\pm$0.3   & 0.002$\pm$0.001     & 7$\pm$2   \\  
   & 40 & 11.3$\pm$0.6 & -15.2$\pm$0.3 & 0.112$\pm$0.008 & 1.6$\pm$0.2   & 40 &  2.3$\pm$0.7 &  -16.2$\pm$0.3   & 0.002$\pm$0.001     & 9$\pm$2   \\  
   & 50 & 11.4$\pm$0.6 & -15.2$\pm$0.3 & 0.087$\pm$0.006 & 1.9$\pm$0.3   & 50 &  2.3$\pm$0.7 &  -16.2$\pm$0.3  & 0.0013$\pm$0.0005   & 11$\pm$3   \\
P2a & 20 & 10.6$\pm$0.7 & -14.7$\pm$0.3 & 0.25$\pm$0.02   & 0.9$\pm$0.1   & 20 &  1.8$\pm$0.3 &  -15.6$\pm$0.1   & 0.0027$\pm$0.0005   & 5$\pm$1   \\  
   & 30 & 10.8$\pm$0.7 & -14.7$\pm$0.3 & 0.15$\pm$0.02   & 1.2$\pm$0.2   & 30 &  1.8$\pm$0.3 &  -15.6$\pm$0.1   & 0.0017$\pm$0.0003   & 7$\pm$1   \\  
   & 40 & 10.9$\pm$0.7 & -14.7$\pm$0.3 & 0.105$\pm$0.008 & 1.4$\pm$0.2   & 40 &  1.8$\pm$0.3 &  -15.6$\pm$0.1   & 0.0013$\pm$0.0003   & 9$\pm$2   \\  
   & 50 & 10.9$\pm$0.7 & -14.7$\pm$0.3 & 0.082$\pm$0.006 & 1.7$\pm$0.3   & 50 &  1.8$\pm$0.3 &  -15.6$\pm$0.1   & 0.0010$\pm$0.0002   & 11$\pm$2   \\
P3a & 20 & 11$\pm$1     & -15.0$\pm$0.5 & 0.17$\pm$0.02   & 0.6$\pm$0.1   & 20 &  2.3$\pm$0.6 &  -16.6$\pm$0.2   & 0.002$\pm$0.001     & 5$\pm$1   \\  
   & 30 & 11$\pm$1     & -15.0$\pm$0.5 & 0.10$\pm$0.02   & 0.8$\pm$0.1   & 30 &  2.3$\pm$0.6 &  -16.6$\pm$0.2   & 0.0013$\pm$0.0003   & 7$\pm$1   \\  
   & 40 & 11$\pm$1     & -15.0$\pm$0.5 & 0.073$\pm$0.009 & 1.0$\pm$0.2   & 40 &  2.3$\pm$0.6 &  -16.6$\pm$0.2   & 0.0010$\pm$0.0002   & 9$\pm$2   \\  
   & 50 & 11$\pm$1     & -15.0$\pm$0.5 & 0.057$\pm$0.007 & 1.2$\pm$0.2   & 50 &  2.3$\pm$0.6 &  -16.6$\pm$0.2   & 0.0008$\pm$0.0002   & 11$\pm$2   \\
P4a & 20 & 10.2$\pm$0.6 & -14.6$\pm$0.3 & 0.20$\pm$0.02   & 0.7$\pm$0.1   & 20 &  2.7$\pm$0.6 &  -15.6$\pm$0.2   & 0.0021$\pm$0.0005   & 6$\pm$1   \\  
   & 30 & 10.3$\pm$0.6 & -14.6$\pm$0.3 & 0.12$\pm$0.01   & 0.9$\pm$0.1   & 30 &  2.7$\pm$0.6 &  -15.6$\pm$0.2   & 0.0013$\pm$0.0003   & 8$\pm$2   \\  
   & 40 & 10.3$\pm$0.6 & -14.6$\pm$0.3 & 0.089$\pm$0.007 & 1.2$\pm$0.2   & 40 &  2.7$\pm$0.6 &  -15.6$\pm$0.2   & 0.0009$\pm$0.0002   & 10$\pm$2   \\
   & 50 & 10.4$\pm$0.6 & -14.6$\pm$0.3 & 0.069$\pm$0.005 & 1.4$\pm$0.2   & 50 &  2.7$\pm$0.6 &  -15.6$\pm$0.2   & 0.0007$\pm$0.0002   & 13$\pm$3   \\
   \hline  \hline
   Source & \multicolumn{5}{c}{\NN(1--0)} & \multicolumn{5}{c}{\N15(1--0)} \\
  & $T_{\rm ex}$  & $FWHM^{a}$  &$v_{\rm LSR}$ & $\tau$& $N_{\rm tot}$ & $T_{\rm ex}$  & $FWHM$ &$v_{\rm LSR}$ & $\tau^{b}$ & $N_{\rm tot}$ \\
 & (K) &(km s$^{-1}$) & (km s$^{-1}$) &  & ($\times$10$^{14}$cm$^{-2}$)  & (K) &(km s$^{-1}$) & (km s$^{-1}$)  & & ($\times$10$^{11}$cm$^{-2}$) \\
 \hline
P1b & 20 & 11.2$\pm$0.7 & -15.2$\pm$0.3 & 0.27$\pm$0.02   & 1.0$\pm$0.2   & 20 &  1.2$\pm$0.2 &  -16.27$\pm$0.07 & 0.004$\pm$0.001     & 5$\pm$1    \\   
   & 30 & 11.4$\pm$0.7 & -15.2$\pm$0.3 & 0.16$\pm$0.02   & 1.3$\pm$0.2   & 30 &  1.2$\pm$0.2 &  -16.27$\pm$0.07 & 0.0022$\pm$0.0004   & 6$\pm$1    \\ 
   & 40 & 11.5$\pm$0.7 & -15.2$\pm$0.3 & 0.114$\pm$0.009 & 1.7$\pm$0.3   & 40 &  1.2$\pm$0.2 &  -16.27$\pm$0.07 & 0.0016$\pm$0.0003   & 8$\pm$2    \\ 
   & 50 & 11.5$\pm$0.7 & -15.2$\pm$0.3 & 0.089$\pm$0.007 & 2.0$\pm$0.3   & 50 &  1.2$\pm$0.2 &  -16.27$\pm$0.07 & 0.0013$\pm$0.0002   & 9$\pm$2    \\ 
P2b & 20 & 11.0$\pm$0.7 & -14.7$\pm$0.3 & 0.23$\pm$0.02   & 0.9$\pm$0.1   & 20 &  1.9$\pm$0.3 &  -15.5$\pm$0.1   & 0.004$\pm$0.001     & 7$\pm$2    \\   
   & 30 & 11.1$\pm$0.6 & -14.7$\pm$0.3 & 0.14$\pm$0.01   & 1.2$\pm$0.2   & 30 &  1.9$\pm$0.3 &  -15.5$\pm$0.1   & 0.0023$\pm$0.0004   & 10$\pm$2   \\
   & 40 & 11.2$\pm$0.6 & -14.7$\pm$0.3 & 0.110$\pm$0.007 & 1.4$\pm$0.2   & 40 &  1.9$\pm$0.3 &  -15.5$\pm$0.1   & 0.0016$\pm$0.0003   & 13$\pm$3   \\
   & 50 & 11.2$\pm$0.6 & -14.7$\pm$0.3 & 0.078$\pm$0.006 & 1.7$\pm$0.3   & 50 &  1.9$\pm$0.3 &  -15.5$\pm$0.1   & 0.0013$\pm$0.0002   & 15$\pm$3   \\
P3b & 20 & 11$\pm$1     & -14.9$\pm$0.5 & 0.20$\pm$0.03   & 0.7$\pm$0.1   & 20 &  2.0$\pm$0.3 &  -16.5$\pm$0.2   & 0.003$\pm$0.001     & 7$\pm$1    \\   
   & 30 & 11$\pm$1     & -14.9$\pm$0.5 & 0.12$\pm$0.02   & 0.9$\pm$0.2   & 30 &  2.0$\pm$0.3 &  -16.5$\pm$0.2   & 0.0019$\pm$0.0004   & 9$\pm$2    \\   
   & 40 & 11$\pm$1     & -14.9$\pm$0.5 & 0.08$\pm$0.01   & 1.2$\pm$0.2   & 40 &  2.0$\pm$0.3 &  -16.5$\pm$0.2   & 0.0014$\pm$0.0003   & 11$\pm$3   \\
   & 50 & 11$\pm$1     & -14.9$\pm$0.5 & 0.065$\pm$0.008 & 1.4$\pm$0.3   & 50 &  2.0$\pm$0.3 &  -16.5$\pm$0.2   & 0.0011$\pm$0.0002   & 14$\pm$3   \\
P4b & 20 & 10.7$\pm$0.6 & -14.7$\pm$0.3 & 0.082$\pm$0.006 & 0.30$\pm$0.04 & 20 &  1.5$\pm$0.3 &  -16.7$\pm$0.2   & 0.0010$\pm$0.0003   & 1.6$\pm$0.4   \\
   & 30 & 10.7$\pm$0.6 & -14.7$\pm$0.3 & 0.051$\pm$0.004 & 0.40$\pm$0.06 & 30 &  1.5$\pm$0.3 &  -16.7$\pm$0.2   & 0.0006$\pm$0.0002   & 2.2$\pm$0.6   \\
   & 40 & 10.7$\pm$0.6 & -14.7$\pm$0.3 & 0.037$\pm$0.003 & 0.50$\pm$0.07 & 40 &  1.5$\pm$0.3 &  -16.7$\pm$0.2   & 0.0005$\pm$0.0001   & 2.7$\pm$0.7   \\
   & 50 & 10.7$\pm$0.6 & -14.7$\pm$0.3 & 0.029$\pm$0.002 & 0.61$\pm$0.09 & 50 &  1.5$\pm$0.3 &  -16.7$\pm$0.2   & 0.00040$\pm$0.00009 & 3.3$\pm$0.9   \\
  \bottomrule
 \end{longtable}
\begin{flushleft}
 \footnotesize
  $^{a}$ Note that these FWHM overestimate those obtained with the $^{15}$N-isotopologues since \NN has a spectral resolution higher than the physical FWHM;\\
 $^{b}$ opacity of the main hyperfine component.\\
  \end{flushleft}
       \normalsize 
\newpage

\begin{longtable}{l*{10}{c}}
\caption{\label{inters-fit}Values obtained with the fitting procedure described in Sect.~\ref{madcuba} to the \NN(1--0), \15N(1--0) and \N15(1--0) transitions towards I1, I2, I3 and I4.}
\tabularnewline \toprule \cline{1-6}  
Source & \multicolumn{5}{c}{\NN(1--0)} & & & & & \\
  & $T_{\rm ex}$  & $FWHM^{a}$ &$v_{\rm LSR}$ & $\tau$  & $N_{\rm tot}$ & & & & & \\
 & (K) &(km s$^{-1}$) & (km s$^{-1}$) &  & ($\times$10$^{14}$cm$^{-2}$)  & & & & & \\
 \cline{1-6}
  \endhead
  I1 & 20 & 11.3$\pm$0.7 & -15.3$\pm$0.3 & 0.29$\pm$0.03   & 1.1$\pm$0.2 & & & & & \\
   & 30 & 11.5$\pm$0.7 & -15.3$\pm$0.3 & 0.17$\pm$0.02   & 1.5$\pm$0.2 & & & & & \\
   & 40 & 11.6$\pm$0.7 & -15.3$\pm$0.3 & 0.124$\pm$0.009 & 1.8$\pm$0.3 & & & & & \\
   & 50 & 11.7$\pm$0.7 & -15.3$\pm$0.3 & 0.096$\pm$0.007 & 2.2$\pm$0.3 & & & & & \\
I2 & 20 & 11.0$\pm$0.6 & -14.7$\pm$0.3 & 0.25$\pm$0.02   & 1.0$\pm$0.2 & & & & & \\
   & 30 & 11.1$\pm$0.7 & -14.7$\pm$0.3 & 0.15$\pm$0.02   & 1.2$\pm$0.2 & & & & & \\
   & 40 & 11.2$\pm$0.7 & -14.7$\pm$0.3 & 0.109$\pm$0.009 & 1.5$\pm$0.3 & & & & & \\
   & 50 & 11.3$\pm$0.7 & -14.7$\pm$0.3 & 0.084$\pm$0.007 & 1.9$\pm$0.3 & & & & & \\
I3 & 20 & 11.0$\pm$0.9 & -15.0$\pm$0.5 & 0.17$\pm$0.02   & 0.6$\pm$0.1 & & & & & \\ 
   & 30 & 11.1$\pm$0.9 & -15.0$\pm$0.5 & 0.10$\pm$0.01   & 0.8$\pm$0.2 & & & & & \\ 
   & 40 & 11.2$\pm$0.9 & -15.0$\pm$0.5 & 0.073$\pm$0.008 & 1.0$\pm$0.2 & & & & & \\
   & 50 & 11.2$\pm$0.9 & -15.0$\pm$0.5 & 0.057$\pm$0.006 & 1.2$\pm$0.2 & & & & & \\
I4 & 20 & 10.8$\pm$0.6 & -14.8$\pm$0.3 & 0.18$\pm$0.02   & 0.7$\pm$0.1 & & & & & \\ 
   & 30 & 10.9$\pm$0.6 & -14.8$\pm$0.3 & 0.111$\pm$0.008 & 0.9$\pm$0.1 & & & & & \\ 
   & 40 & 11.0$\pm$0.6 & -14.7$\pm$0.3 & 0.079$\pm$0.006 & 1.1$\pm$0.2 & & & & & \\
   & 50 & 11.0$\pm$0.6 & -14.7$\pm$0.3 & 0.062$\pm$0.004 & 1.3$\pm$0.2 & & & & & \\
   \hline  \hline
   Source & \multicolumn{5}{c}{\15N(1--0)} & \multicolumn{5}{c}{\N15(1--0)} \\
  & $T_{\rm ex}$  & $FWHM$ &$v_{\rm LSR}$ & $\tau^{b}$& $N_{\rm tot}$ & $T_{\rm ex}$  & $FWHM$ &$v_{\rm LSR}$ & $\tau^{b}$ & $N_{\rm tot}$ \\
 & (K) &(km s$^{-1}$) & (km s$^{-1}$) &  & ($\times$10$^{11}$cm$^{-2}$)  & (K) &(km s$^{-1}$) & (km s$^{-1}$)  & & ($\times$10$^{11}$cm$^{-2}$) \\
 \hline
I1 & 20 & 1.7$\pm$0.5 & -16.4$\pm$0.2 & 0.005$\pm$0.001   & 5$\pm$1   & 20 & 1.3$\pm$0.3 & -16.1$\pm$0.1 & 0.005$\pm$0.001   & 4$\pm$1    \\
   & 30 & 1.7$\pm$0.5 & -16.4$\pm$0.2 & 0.002$\pm$0.001   & 7$\pm$2   & 30 & 1.3$\pm$0.3 & -16.1$\pm$0.1 & 0.003$\pm$0.001   & 6$\pm$2    \\ 
   & 40 & 1.7$\pm$0.5 & -16.4$\pm$0.2 & 0.002$\pm$0.001   & 9$\pm$2   & 40 & 1.3$\pm$0.3 & -16.1$\pm$0.1 & 0.002$\pm$0.001   & 7$\pm$2    \\ 
   & 50 & 1.7$\pm$0.5 & -16.4$\pm$0.2 & 0.002$\pm$0.001   & 11$\pm$2 & 50 & 1.3$\pm$0.3 & -16.1$\pm$0.1 & 0.0019$\pm$0.0005  & 9$\pm$2    \\
I2 & 20 & 1.6$\pm$0.4 & -15.7$\pm$0.2 & 0.006$\pm$0.002   & 6$\pm$1   & 20 & 1.7$\pm$0.3 & -15.5$\pm$0.1 & 0.007$\pm$0.001   & 7$\pm$2    \\
   & 30 & 1.9$\pm$0.3 & -15.6$\pm$0.1 & 0.0030$\pm$0.0005 & 8$\pm$2   & 30 & 1.7$\pm$0.3 & -15.5$\pm$0.1 & 0.004$\pm$0.001   & 10$\pm$2  \\
   & 40 & 1.9$\pm$0.3 & -15.6$\pm$0.1 & 0.0022$\pm$0.0003 & 11$\pm$2 & 40 & 1.7$\pm$0.3 & -15.5$\pm$0.1 & 0.003$\pm$0.001    & 12$\pm$3  \\ 
   & 50 & 1.9$\pm$0.3 & -15.6$\pm$0.1 & 0.0017$\pm$0.0003 & 13$\pm$2 & 50 & 1.7$\pm$0.3 & -15.5$\pm$0.1 & 0.0024$\pm$0.0004  & 15$\pm$3  \\
I3 & 20 & 2.6$\pm$0.6 & -16.6$\pm$0.2 & 0.003$\pm$0.001   & 6$\pm$1   & 20 & 2.2$\pm$0.3 & -16.5$\pm$0.2 & 0.005$\pm$0.001   & 7$\pm$2    \\
   & 30 & 2.6$\pm$0.6 & -16.6$\pm$0.2 & 0.002$\pm$0.001   & 7$\pm$2   & 30 & 2.2$\pm$0.3 & -16.5$\pm$0.2 & 0.003$\pm$0.001   & 9$\pm$2    \\
   & 40 & 2.6$\pm$0.6 & -16.6$\pm$0.2 & 0.0015$\pm$0.0004 & 9$\pm$2   & 40 & 2.2$\pm$0.3 & -16.5$\pm$0.2 & 0.0021$\pm$0.0004 & 11$\pm$3  \\
   & 50 & 2.6$\pm$0.6 & -16.6$\pm$0.2 & 0.0012$\pm$0.0003 & 11$\pm$2 & 50 & 2.2$\pm$0.3 & -16.5$\pm$0.2 & 0.0017$\pm$0.0003  & 14$\pm$3  \\
I4 & 20 & 3.2$\pm$0.8 & -15.3$\pm$0.3 & 0.003$\pm$0.001   & 7$\pm$2   & 20 & 1.4$\pm$0.3 & -16.4$\pm$0.1 & 0.006$\pm$0.001   & 5$\pm$1    \\
   & 30 & 3.2$\pm$0.8 & -15.3$\pm$0.3 & 0.002$\pm$0.001   & 9$\pm$2   & 30 & 1.4$\pm$0.3 & -16.4$\pm$0.1 & 0.003$\pm$0.001   & 6$\pm$2    \\
   & 40 & 3.2$\pm$0.8 & -15.3$\pm$0.3 & 0.0014$\pm$0.0004 & 11$\pm$3 & 40 & 1.4$\pm$0.3 & -16.4$\pm$0.1 & 0.003$\pm$0.001    & 8$\pm$2    \\
   & 50 & 3.2$\pm$0.8 & -15.3$\pm$0.3 & 0.0011$\pm$0.0003 & 14$\pm$3 & 50 & 1.4$\pm$0.3 & -16.4$\pm$0.1 & 0.002$\pm$0.001    & 10$\pm$3  \\
  \bottomrule
 \end{longtable}
\begin{flushleft}
 \footnotesize
 $^{a}$ Note that these FWHM overestimate those obtained with the $^{15}$N-isotopologues since \NN has a spectral resolution higher than the physical FWHM;\\
 $^{b}$ opacity of the main hyperfine component.\\
   \end{flushleft}
       \normalsize  
\newpage

\begin{longtable}{l*{10}{c}}
\caption{\label{diffuse-fit}Values obtained with the fitting procedure described in Sect.~\ref{madcuba} to the \NN(1--0), \15N(1--0) and \N15(1--0) transitions towards D1, D2 and D3. The upper limits and tentative detections are obtained as explained in Sect.~\ref{diffuse}.}
\tabularnewline \toprule \cline{1-6}  
Source & \multicolumn{5}{c}{\NN(1--0)} & & & & & \\
  & $T_{\rm ex}$  & $FWHM^{a}$ &$v_{\rm LSR}$ & $\tau$  & $N_{\rm tot}$ & & & & & \\
 & (K) &(km s$^{-1}$) & (km s$^{-1}$) &  & ($\times$10$^{13}$cm$^{-2}$)  & & & & & \\
 \cline{1-6}
  \endhead
 D1 & 20 & 10.0$\pm$0.7 & -15.0$\pm$0.4 & 0.14$\pm$0.02     & 4.9$\pm$0.8  & & & & & \\
   & 30 & 10.0$\pm$0.7 & -15.0$\pm$0.4 & 0.087$\pm$0.008   & 6$\pm$1      & & & & & \\
   & 40 & 10.1$\pm$0.7 & -15.0$\pm$0.4 & 0.063$\pm$0.006   & 8$\pm$2      & & & & & \\
   & 50 & 10.1$\pm$0.7 & -15.0$\pm$0.4 & 0.049$\pm$0.004   & 10$\pm$2     & & & & & \\
   \cline{1-6}
D2 & 20 & 8.9$\pm$0.9  & -14.4$\pm$0.4 & 0.12$\pm$0.02     & 3.7$\pm$0.7  &  \\
   & 30 & 9.0$\pm$0.9  & -14.4$\pm$0.4 & 0.075$\pm$0.009   & 4.9$\pm$0.9  &  \\ 
   & 40 & 9.0$\pm$0.9  & -14.4$\pm$0.4 & 0.054$\pm$0.007   & 6$\pm$1      &  \\ 
   & 50 & 9.0$\pm$0.9  & -14.4$\pm$0.4 & 0.042$\pm$0.005   & 7$\pm$2      &  \\
   \cline{1-6}
D3 & 20 & 9$\pm$1      & -14.8$\pm$0.5 & 0.11$\pm$0.02     & 3.4$\pm$0.7  & & & & & \\
   & 30 & 9$\pm$1      & -14.7$\pm$0.5 & 0.069$\pm$0.01    & 4.5$\pm$0.9  & & & & & \\
   & 40 & 9$\pm$1      & -14.7$\pm$0.5 & 0.050$\pm$0.007   & 6$\pm$1      & & & & & \\
   & 50 & 9$\pm$1      & -14.7$\pm$0.5 & 0.039$\pm$0.006   & 7$\pm$2      & & & & & \\
    \hline  \hline
   Source & \multicolumn{5}{c}{\15N(1--0)} & \multicolumn{5}{c}{\N15(1--0)} \\
  & $T_{\rm ex}$  & $FWHM$ &$v_{\rm LSR}$ & $\tau^{b}$& $N_{\rm tot}$ & $T_{\rm ex}$  & $FWHM$ &$v_{\rm LSR}$ & $\tau^{b}$ & $N_{\rm tot}$ \\
 & (K) &(km s$^{-1}$) & (km s$^{-1}$) &  & ($\times$10$^{11}$cm$^{-2}$)  & (K) &(km s$^{-1}$) & (km s$^{-1}$)  & & ($\times$10$^{11}$cm$^{-2}$) \\
 \hline
D1 & 20 &    & & & $\leq$2   & 20 &    & & & $\leq$2.4 \\
   & 30 &    & & & $\leq$2.6 & 30 &    & & & $\leq$3.2 \\
   & 40 &    & & & $\leq$3.3 & 40 &    & & & $\leq$4   \\
   & 50 &    & & & $\leq$4   & 50 &    & & & $\leq$4.9 \\
   \hline
D2 & 20 & 0.8$\pm$0.2 & -15.1$\pm$0.1 & 0.002$\pm$0.001   & 1.1$\pm$0.4$^{c}$ & 20 &   & & & $\leq$2.4 \\
   & 30 & 0.8$\pm$0.2 & -15.1$\pm$0.1 & 0.0014$\pm$0.0005 & 1.5$\pm$0.6$^{c}$ & 30 &    & & & $\leq$3.3 \\
   & 40 & 0.8$\pm$0.2 & -15.1$\pm$0.1 & 0.0010$\pm$0.0004 & 1.9$\pm$0.7$^{c}$ & 40 &    & & & $\leq$4.2 \\
   & 50 & 0.8$\pm$0.2 & -15.1$\pm$0.1 & 0.0008$\pm$0.0003 & 2.4$\pm$0.8$^{c}$ & 50 &    & & & $\leq$5   \\
   \hline
D3 & 20 &    & & & $\leq$1.4 & 20 &    & & & $\leq$1   \\
   & 30 &    & & & $\leq$1.8 & 30 &    & & & $\leq$1.4 \\
   & 40 &    & & & $\leq$2.3 & 40 &    & & & $\leq$1.7 \\
   & 50 &    & & & $\leq$2.8 & 50 &    & & & $\leq$2.1 \\
  \bottomrule
 \end{longtable}
 \begin{flushleft}
 \footnotesize
  $^{a}$ Note that these FWHM overestimate those obtained with the $^{15}$N-isotopologues since \NN has a spectral resolution higher than the physical FWHM;\\
 $^{b}$ opacity of the main hyperfine component;\\
  $^{c}$ tentative detection.
   \end{flushleft}
       \normalsize  
  
  \begin{longtable}{l*{10}{c}}
\caption{\label{30m-fit}Values obtained with the fitting procedure described in Sect.~\ref{madcuba} to the \NN(1--0), \15N(1--0) and \N15(1--0) transitions towards mm1 and mm3 in regions equals to the IRAM-30m beam (see Sect~\ref{extract-30m}).}
\tabularnewline \toprule \cline{1-6}  
Source & \multicolumn{5}{c}{\NN(1--0)} & & & & & \\
  & $T_{\rm ex}$  & $FWHM^{a}$ &$v_{\rm LSR}$ & $\tau$  & $N_{\rm tot}$ & & & & & \\
 & (K) &(km s$^{-1}$) & (km s$^{-1}$) &  & ($\times$10$^{13}$cm$^{-2}$)  & & & & & \\
 \cline{1-6}
  \endhead
 mm1 & 43 & 10.0$\pm$0.7 & -14.8$\pm$0.4  & 0.061$\pm$0.006   & 9$\pm$2     & & & & & \\
mm3 & 18 & 9.9$\pm$0.6  & -14.7$\pm$0.3  & 0.18$\pm$0.02     & 5.0$\pm$0.8 & & & & & \\
    \hline  \hline
   Source & \multicolumn{5}{c}{\15N(1--0)} & \multicolumn{5}{c}{\N15(1--0)} \\
  & $T_{\rm ex}$  & $FWHM$ &$v_{\rm LSR}$ & $\tau^{b}$& $N_{\rm tot}$ & $T_{\rm ex}$  & $FWHM$ &$v_{\rm LSR}$ & $\tau^{b}$ & $N_{\rm tot}$ \\
 & (K) &(km s$^{-1}$) & (km s$^{-1}$) &  & ($\times$10$^{11}$cm$^{-2}$)  & (K) &(km s$^{-1}$) & (km s$^{-1}$)  & & ($\times$10$^{11}$cm$^{-2}$) \\
 \hline
mm1 & 43 & 1.7$\pm$0.3  & -16.1$\pm$0.1  & 0.0009$\pm$0.0001 & 4.1$\pm$0.7 & 43 & 2.1$\pm$0.4 & -16.2$\pm$0.2 & 0.0008$\pm$0.0002 & 5$\pm$2     \\
mm3 & 18 & 1.9$\pm$0.4  & -15.8$\pm$0.1  & 0.0022$\pm$0.0004 & 2.3$\pm$0.4 & 18 & 1.8$\pm$0.3 & -16.1$\pm$0.2 & 0.003$\pm$0.001   & 2.6$\pm$0.7 \\  
  \bottomrule
 \end{longtable}
\begin{flushleft}
 \footnotesize
  $^{a}$ Note that these FWHM overestimate those obtained with the $^{15}$N-isotopologues since \NN has a spectral resolution higher than the physical FWHM;\\
 $^{b}$ opacity of the main hyperfine component;\\
   \end{flushleft}
       \normalsize

\end{document}